\documentclass[12pt]{article}
\pdfoutput=1
\usepackage{graphicx}
\usepackage{amsmath,amssymb,multirow,xspace}
\usepackage[colorlinks=true,urlcolor=blue,anchorcolor=blue,citecolor=blue,filecolor=blue,linkcolor=blue,menucolor=blue,pagecolor=blue]{hyperref}
\usepackage[compress,numbers]{natbib}
\usepackage{caption}
\usepackage{subcaption}
\usepackage{soul}
\usepackage{placeins}
\usepackage{booktabs}
\usepackage{blindtext, rotating}
\usepackage{afterpage}
\usepackage{enumitem}
\usepackage{marvosym }
\usepackage{verbatim}
\usepackage{placeins}
\usepackage{pbox}
\usepackage{slashed}
\usepackage{multirow}
\usepackage{floatrow}
\newfloatcommand{capbtabbox}{table}[][\FBwidth]

\usepackage[font=footnotesize,labelfont=bf]{caption}
\usepackage[font=footnotesize,labelfont=bf]{subcaption}

\long\def\symbolfootnote[#1]#2{\begingroup%
\def\thefootnote{\fnsymbol{footnote}}\footnote[#1]{#2}\endgroup}

\newcommand{\newc}{\newcommand}
\newc{\gsim}{\lower.7ex\hbox{$\;\stackrel{\textstyle>}{\sim}\;$}}
\newc{\lsim}{\lower.7ex\hbox{$\;\stackrel{\textstyle<}{\sim}\;$}}
\newc{\gev}{\,{\rm GeV}}
\newc{\mev}{\,{\rm MeV}}
\newc{\ev}{\,{\rm eV}}
\newc{\kev}{\,{\rm keV}}
\newc{\tev}{\,{\rm TeV}}

\newc{\mz}{M_Z}
\newc{\mpl}{M_*}
\newc{\mw}{m_{\rm weak}}
\newc{\nr}[1]{N^c_R{}_{#1}}
\usepackage{amsmath}

\def\beq{\begin{equation}}
\def\eeq{\end{equation}}
\newcommand{\bea}{\begin{eqnarray}\begin{aligned}}
\newcommand{\eea}{\end{aligned}\end{eqnarray}}
\def\bitem{\begin{itemize}}
\def\eitem{\end{itemize}}
\newc{\ie}{{\it i.e.}}          \newc{\etal}{{\it et al.}}
\newc{\eg}{{\it e.g.}}          \newc{\etc}{{\it etc.}}
\newc{\cf}{{\it c.f.}}

 \numberwithin{equation}{section}

\newcommand\fverb{\setbox\fverbbox=\hbox\bgroup\verb}
\newcommand\fverbdo{\egroup\medskip\noindent%
            \fbox{\unhbox\fverbbox}\ }
\newcommand\fverbit{\egroup\item[\fbox{\unhbox\fverbbox}]}
\newbox\fverbbox

\usepackage{authblk}

\usepackage[usenames,dvipsnames,svgnames,table]{xcolor}

\begin{document}

\author[1,2]{Simon Knapen\thanks{smknapen@lbl.gov}}
\author[3,4]{Diego Redigolo\thanks{dredigol@lpthe.jussieu.fr}}

\small
\affil[1]{\small{Berkeley Center for Theoretical Physics,
University of California, Berkeley, CA 94720}
}
\affil[2]{ Theoretical Physics Group, Lawrence Berkeley National Laboratory, Berkeley, CA 94720
}
\affil[3]{Sorbonne Universit\'es, UPMC Univ Paris 06, UMR 7589, LPTHE, F-75005, Paris, France
}
\affil[4]{CNRS, UMR 7589, LPTHE, F-75005, Paris, France
}

\title{Gauge Mediation at the LHC: status and prospects}

\maketitle
\begin{abstract}
We show that the predictivity of general gauge mediation (GGM) with TeV-scale stops is greatly increased once the Higgs mass constraint is imposed. The most notable results are a strong lower bound on the mass of the gluino and right-handed squarks, and an upper bound on the Higgsino mass. If the $\mu$-parameter is positive, the wino mass is also bounded from above.  These constraints relax significantly for high messenger scales and as such long-lived NLSPs are favored in GGM. We identify a small set of most promising topologies for the neutralino/sneutrino NLSP scenarios and estimate the impact of the current bounds and the sensitivity of the high luminosity LHC. The stau, stop and sbottom NLSP scenarios can be robustly excluded at the high luminosity LHC.

\end{abstract}

\clearpage

\tableofcontents

\section{Introduction}
With the early phase of run II of the LHC under way, it is a particularly good time to review what run I has taught us about weak-scale supersymmetry, and to look ahead about what we can expect to learn from the next data set. The most significant lesson from run I was without any doubt the discovery of a SM-like Higgs with a mass of 125 GeV \cite{Aad:2012tfa,Chatrchyan:2012xdj}. This has profound implications on the spectrum of the superpartners, as it tends to favor a somewhat higher supersymmetry breaking scale. In the light of this observation, it is perhaps not so surprising that no concrete evidence for supersymmetry has emerged from the 8 TeV data.

Considering the constraints from precision flavor experiments, manifestly flavor-safe forms of supersymmetry breaking are highly motivated. In this category gauge mediated supersymmetry breaking (GMSB) \cite{Dine:1981za,Dimopoulos:1981au,Dine:1981gu,Dine:1982zb,Nappi:1982hm,AlvarezGaume:1981wy,Dine:1993yw,Dine:1994vc,Dine:1995ag} is perhaps the most elegant and certainly the most widely studied paradigm. In previous work \cite{Knapen:2015qba} we obtained a detailed quantitative understanding of the role of the Higgs mass constraint in the framework of General Gauge Mediation (GGM), which encompasses a wide class of GMSB models \cite{Meade:2008wd,Buican:2008ws}. In particular, we showed that it is highly non-trivial within GGM to simultaneously achieve electroweak symmetry breaking and a sufficiently large $A$-term to accommodate a 125 GeV Higgs with LHC-accessible stops.  
Although it is possible to relax this tension in various extensions of GGM \cite{Evans:2011bea,Evans:2012hg,Kang:2012ra,Craig:2012xp,Abdullah:2012tq,Kim:2012vz,Byakti:2013ti,Craig:2013wga,Evans:2013kxa,Calibbi:2013mka,Jelinski:2013kta,Galon:2013jba,Fischler:2013tva,Knapen:2013zla,Ding:2013pya,Calibbi:2014yha,Basirnia:2015vga,Jelinski:2015gsa,Jelinski:2015voa,Ierushalmi:2016axs,Casas:2016xnl,Allanach:2015mwa,Allanach:2015cia,Allanach:2016pam}, it is also interesting to take it at face value. In this case requiring $m_h=125$ GeV induces strong correlations in spectrum of the superpartners, as well as interesting lower and/or upper bounds on some of the sparticle masses. As a result, the GGM framework is surprisingly predictive, even though it contains as many as 8 independent parameters. 

In the present paper we make use of the understanding of the GGM parameter space developed in \cite{Knapen:2015qba} to survey the phenomenology and to investigate the existing constraints and discovery prospects for GGM at the LHC. While our previous paper was primarily intended for an audience of theorists, we hope this paper will be of some interest to experimentalists as well. 
A general lesson is that accounting for the Higgs mass constraint systematically in top-down SUSY scenarios can drastically modify our expectations for the ``first  signatures'' of SUSY at hadron colliders. This makes full phenomenological surveys of UV motivated scenarios complementary to the bottom up approach based on the full coverage of SUSY decay topologies in terms of simplified models \cite{Alves:2011wf,Cohen:2013xda}. (For simplified models specific to gauge mediation, see \cite{Dimopoulos:1996vz,Dimopoulos:1996fj,Meade:2009qv,Meade:2010ji,Ruderman:2010kj,Ruderman:2011vv,Kats:2011it,Kats:2011qh,Heisig:2015yla}.) For instance we will see how the GGM parameter space, with the Higgs mass accounted for, features strong correlations in the spectrum which single out a small set of particularly promising decay topologies.  

While the full MSSM parameter space is very rich and complicated, we can get a sense of the broader picture by dissecting its phenomenology in terms of the parameters controlling the colored production. These are the squark masses, here parametrized by the geometric mean of the stop masses $M_{s}=\sqrt{m_{\tilde{t}_1}m_{\tilde{t}_2}}$, and the gluino soft mass $M_3$. (This (over)simplified picture is somewhat analogous to the $m_0$ and $M_{1/2}$ parametrization in the mSUGRA framework and relies on the fact that first and second generation squarks masses are tightly correlated with the stop masses.)

It is possible that the Higgs is as heavy as $125$ GeV mostly because the scalars, and in particular the stops, are heavier than 5 TeV. In this case the squarks are clearly inaccessible at the LHC, while the gluino may or may not be accessible. The case with accessible gluinos is often referred to as (mini-)split supersymmetry \cite{Giudice:2004tc,ArkaniHamed:2004fb,ArkaniHamed:2004yi,ArkaniHamed:2012gw,Arvanitaki:2012ps,Cohen:2015lyp}, and may produce interesting signatures in the form of prompt or delayed gluino decays. If on the other hand, the gauginos are as heavy as the squarks, there is little hope for hints of SUSY at the LHC. The latter scenario includes minimal gauge mediation (MGM), as recently discussed in \cite{Vega:2015fna}.

  In this paper we focus on the collider phenomenology of the remaining part of the parameter space, where the stops are below 5 TeV. This part of parameter space is complementary to the more standard split-SUSY and heavy SUSY scenarios, and was somehow neglected in previous studies of the GGM phenomenology \cite{Abel:2009ve,Abel:2010vba,Rajaraman:2009ga,Carpenter:2008he,Grajek:2013ola}, essentially because it is difficult to study by taking different 2D slices  in terms of the UV parameters. We estimate the current constraints and the reach of the high luminosity LHC (HL-LHC) by mapping the viable GGM parameter space into a set of representative simplified models and identify the most relevant simplified topologies. For this purpose we focus on the scenario where the NLSP is long-lived on detector time-scales, which is, as we will see, the most generic case once the Higgs mass is imposed.

The remainder of this paper is organized as follows: In section \ref{sec:GGMreview} we first summarize the main results of \cite{Knapen:2015qba} and then extend these results with a discussion of the expected production cross sections and the various NLSP types. In section \ref{sec:neutralNLSP} we focus on the phenomenology of neutral NLSP. We present the most promising decay topologies with their corresponding branching fractions and estimate the current bounds and the projected bounds at HL-LHC.  In section \ref{sec:ch/colNLSP} we discuss the phenomenology of the spectra with a charged or colored NLSP and we summarize our results in section \ref{sec:conclusions}.  Appendix \ref{app:NLSPdecay} contains a brief review of some aspects of the NLSP decay in gauge mediation. We reserve some additional results on squeezed spectra for Appendix \ref{app:squeezed}.

\section{GGM at the weak scale\label{sec:GGMreview}}
\subsection{GGM in a nutshell}\label{sec:GGMnut}
The defining feature of gauge mediation models is that supersymmetry breaking is communicated via a messenger sector to the MSSM by heavy ``messenger'' states which are charged under the SM gauge interactions. Throughout this paper, we will refer to the mass scale of the heavy messengers as the ``messenger scale'' ($M_{mess}$). As a consequence, the MSSM fields feel SUSY-breaking only through SM gauge interactions. Since the SM gauge interactions are manifestly flavor blind, the necessary flavor alignment is achieved trivially.

Without specifying the nature of the messengers sector, it was shown in \cite{Meade:2008wd} that all gauge mediation models can be captured in a single, very predictive equivalence class. This framework goes by the name of ``General Gauge Mediation" (GGM) \cite{Meade:2008wd,Buican:2008ws} and includes all models where the SM gauge interactions are the \emph{only} source of communication between the supersymmetry breaking sector and the MSSM. While this is a very general condition, it nevertheless yields a number of strong predictions regarding the SUSY-breaking parameters:  
\begin{itemize}
\item The Higgsino mass parameter $\mu$ is set by hand, as a parameter independent from the rest of the soft spectrum.
\item The trilinear scalar couplings ($A$-terms) as well as the $B_\mu$ term vanish at the messenger scale $M_{mess}$. 
\item All soft masses are flavor universal at the messenger scale and satisfy the sum rules\footnote{These sum rules are  typically broken in extensions of GGM with extra yukawa-like interactions involving the MSSM Higgs fields \cite{Evans:2011bea,Evans:2012hg,Kang:2012ra,Craig:2012xp,Abdullah:2012tq,Kim:2012vz,Byakti:2013ti,Craig:2013wga,Evans:2013kxa,Calibbi:2013mka,Jelinski:2013kta,Galon:2013jba,Fischler:2013tva,Knapen:2013zla,Ding:2013pya,Calibbi:2014yha,Basirnia:2015vga,Jelinski:2015gsa,Jelinski:2015voa,Ierushalmi:2016axs,Casas:2016xnl,Allanach:2015mwa,Allanach:2015cia,Allanach:2016pam} and also if D-tadpoles are present in the messenger sector \cite{Meade:2008wd,Argurio:2012qt}.}
\bea
& m_{H_u}^2=  m_{H_d}^2= m_{L}^2\\
& m_Q^2-2m_U^2+m_D^2-m_L^2+m_E^2 =0 \\
& 2m_Q^2-m_U^2-m_D^2-2m_L^2+m_E^2=0,
\label{GGMsumrules}
\eea
also at the messenger scale. 
\end{itemize}
Most of these relations receive important corrections under renormalization group (RG) running from the messenger scale down to the weak scale, but they nevertheless leave a strong imprint on the low energy spectrum.
The full GGM parameter space can then be described in terms of 7  parameters plus the messenger scale ($M_{mess}$), which sets the length of the RG-flow. The parametrization we choose here is
\begin{equation}
 M_1,\; M_2,\; M_3,\;  m_{Q}^2,\;  m_{U}^2,\;  m_{L}^2,\; \mu\; \text{ and }\; M_{mess}\, ,\label{UVparameterspace}
\end{equation}
where we take real gaugino masses and $\mu$, but allow
for both positive and negative values all the soft masses and the $\mu$-term. 
In \cite{Buican:2008ws} it was shown that the full parameter space in \eqref{UVparameterspace} can be spanned by explicit models with weakly coupled messengers. In this paper we do not restrict ourselves to a particular model, but instead deal with the complete parameter space in \eqref{UVparameterspace}. 

We will organize our presentation around the following, model independent phenomenological features of GGM:
\begin{itemize}
\item the gravitino is always the LSP\footnote{We neglect the possibility that the MSSM spectrum is sequestered with respect to the SUSY-breaking scale \cite{Craig:2008vs}.} 
\item the NLSP decays to the gravitino LPS and a SM state (which depend on the nature of the NLSP) with a decay width which is suppressed by $M_{mess}$. In appendix~\ref{app:NLSPdecay} we briefly review the main features of the NLSP decay to the gravitino and a SM state.
\end{itemize}
In this paper we focus on NLSP masses around or below the TeV scale (within the reach of LHC) and we take two benchmark datasets for high and low messenger scale gauge mediation, with respectively $M_{mess}=10^{15}\text{ GeV}$ and $M_{mess}=10^{7} \text{ GeV}$. In the former case, we assume that Planck-suppressed contributions from gravity mediation are small compared to those from the gauge mediation sector. (For a discussion of gauge-gravity hybrid models, see for instance \cite{Feng:2007ke,Feng:2009bd}.) As we will see, the phenomenology of these two benchmarks differs greatly.

\subsection{Features of the GGM spectrum}\label{sec:solveGGM}
While the parameter space in \eqref{UVparameterspace} is of course an enormous reduction from the 100+ parameters which characterize the general MSSM, it is still challenging to fully survey this 8 dimensional parameter space. In earlier efforts, this problem has been partially addressed by taking lower dimensional slices \cite{Abel:2009ve,Abel:2010vba,Rajaraman:2009ga,Carpenter:2008he,Grajek:2013ola}, of which the ``Minimal Gauge Mediation'' \cite{Babu:1996jf} slice  is the most well known. While these studies capture a number of generic features, they are insufficient to get a complete picture of the surviving parameter space after the Higgs mass constraint is imposed. In \cite{Knapen:2015qba} we addressed this deficiency by obtaining both a numerical and a semi-analytic solution of the full 8 dimensional parameter space of GGM, including the Higgs mass constraint.\footnote{To account for a theory uncertainty on $m_h$ of a few GeV, we only insist on $m_{h}=123$ GeV, as computed by softsusy-3.5.1 \cite{Allanach:2001kg,Degrassi:2001yf,Brignole:2001jy,Brignole:2002bz,Dedes:2003km}. This ensures that our bounds are conservative.}  Motivated by the Higgs mass, we restrict ourselves to the regime where $\tan\beta \gtrsim$ 10. In this regime the spectrum is fairly insensitive to $\tan\beta$, which we fix to $\tan\beta=20$. Also solving for $m_Z$ and $m_h$ then reduces the parameter space from 8 to 5 dimensions. Of these 5 remaining parameters $M_{mess}$ and $M_1$ can be scanned coarsely without sacrificing a smooth interpolation, such that a manageable 3 dimensional volume remains to be mapped out carefully.

The most important features of the solution are most easily understood in terms of a set of approximate relations between the soft parameters at the weak scale  \cite{Knapen:2015qba}. In the large $\tan\beta$ limit these relations are
\bea
&m^2_{Q_{1,2\phantom{,3}}}\simeq m^2_{Q_3}+{1\over3}(m_{L_3}^2+\mu^2)\\
&m^2_{U_{1,2\phantom{,3}}} \simeq m^2_{U_3}+{2\over 3}(m_{L_3}^2+\mu^2)\\
& m_{L_{1,2\phantom{,3}}}^2\simeq m_{L_3}^2\\
& m_{D_{1,2,3}}^2 \simeq {1\over2}(m_{Q_3}^2+m_{U_3}^2)+{1\over2}\mu^2\\
& m_{E_{1,2,3}}^2 \simeq 2m_{L_3}^2+{1\over2}\mu^2 +  {3\over2}(m_{U_3}^2-m_{Q_3}^2)\\
& m_{H_u}^2\simeq -\mu^2\, \qquad m_{H_d}^2\simeq m_{L_3}^2\, \qquad m_{A^0}^2\simeq m_{L_3}^2+\mu^2.
\label{IRexact} 
\eea
These relations are direct consequences of flavor universality, the electroweak symmetry breaking (EWSB)  conditions and the GGM UV sum-rules in \eqref{GGMsumrules}, and are therefore independent of the messenger scale. A number of phenomenological features immediately follow:
\begin{itemize}
\item 1st/2nd generation Q and U squarks are always heavier of the 3rd generation squarks, although the mass splitting can be small if the left-handed slepton and the Higgsino are both light. 
\item D squarks are always heavier than the quadratic mean of $m_{Q_3}$ and $m_{U_3}$.
\item The right-handed sleptons $E$ are always heavier than the left-handed sleptons $L$~if~$m_{Q_3}<m_{U_3}$. Reversely, for a relatively light Higgsino, the $m_{U_3}\ll m_{Q_3}$ implies that the left-handed sleptons are heavier than the right-handed sleptons.

\item If both the left-handed slepton and the Higgsino are light, so must be the pseudo-scalar Higgs $A^0$. 
\end{itemize}

\begin{figure}[t]\centering
\includegraphics[width=.48\textwidth]{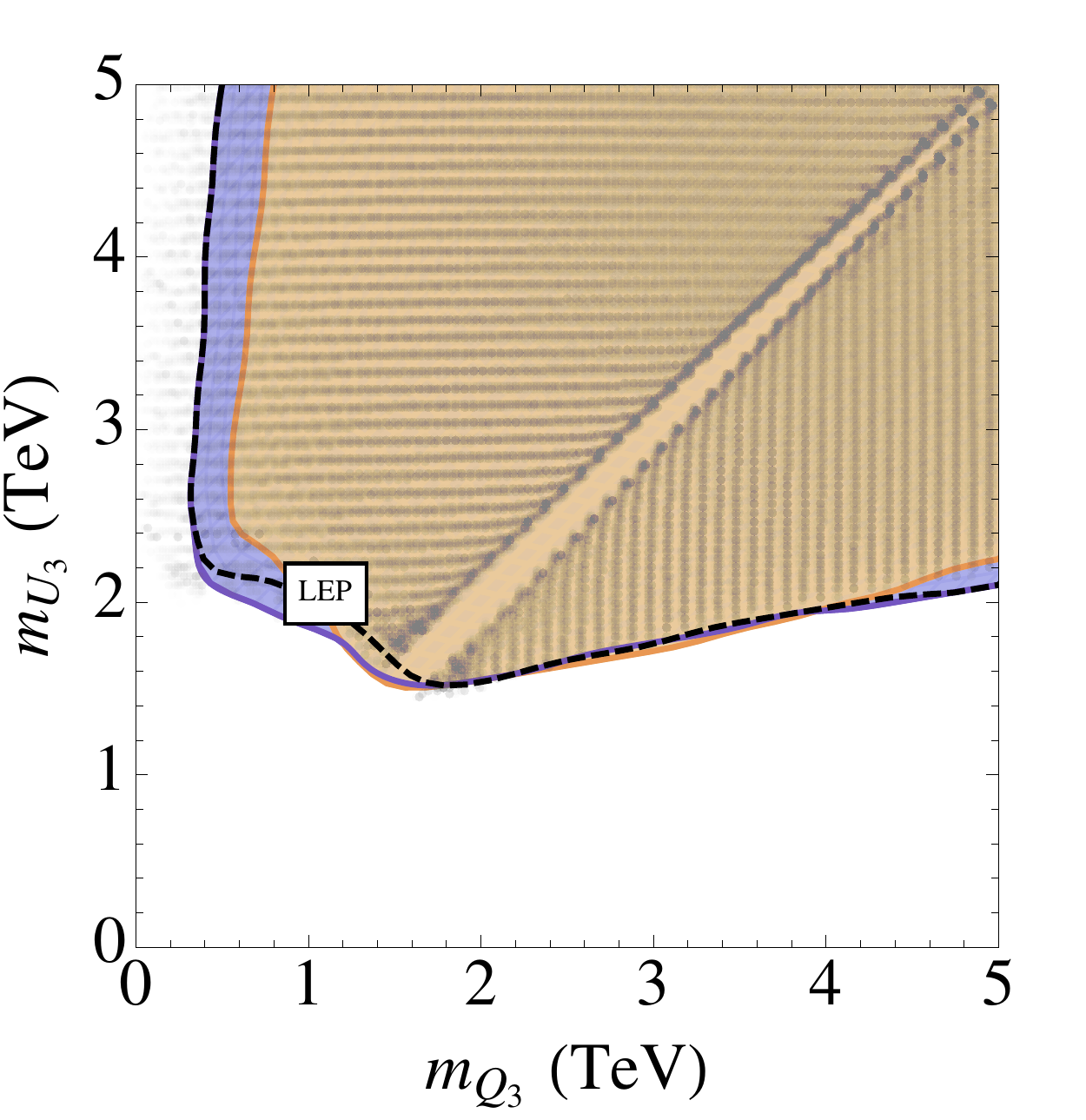}\hfill
\includegraphics[width=.48\textwidth]{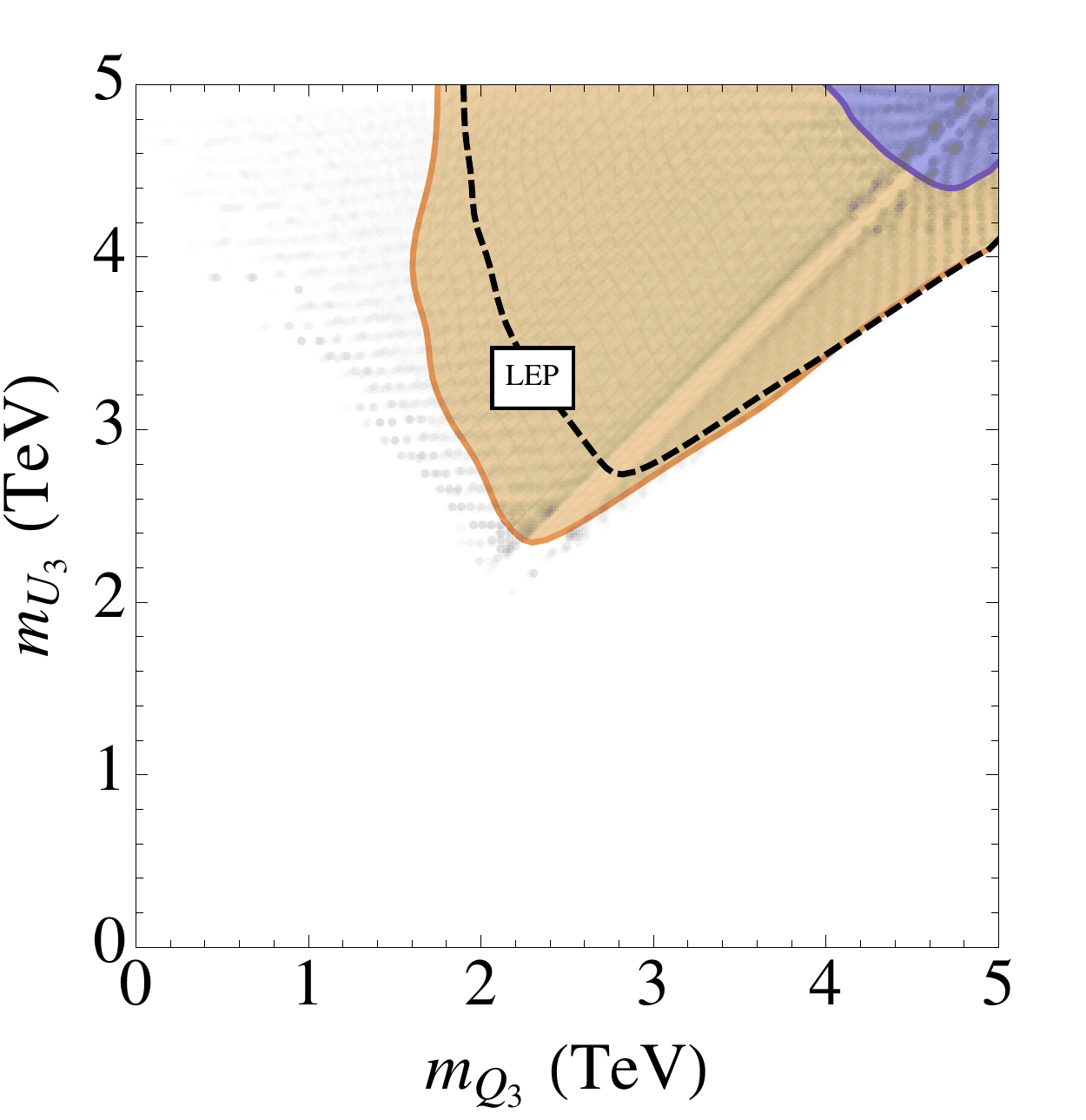}

\caption{Allowed GGM parameter space in the stop soft mass plane with $\mu<0$ ($\mu>0$) corresponding to the blue (orange) shaded regions. The region of the stop mass plane below the dashed line is excluded by LEP bound on the chargino mass. The gray dots show the physical stop masses for the points which passed the LEP bounds. For low $m_Q$,  these are significantly affected by a gluino threshold correction, as explained in the text. Level repulsion between the two stop mass eigenstates causes the wedge along the diagonal. We fixed $M_1=1$ TeV and $\tan\beta=20$ and marginalized over the remaining parameter space. \label{moneyplane}}
\end{figure}

\begin{figure}\centering
\includegraphics[width=0.42\textwidth]{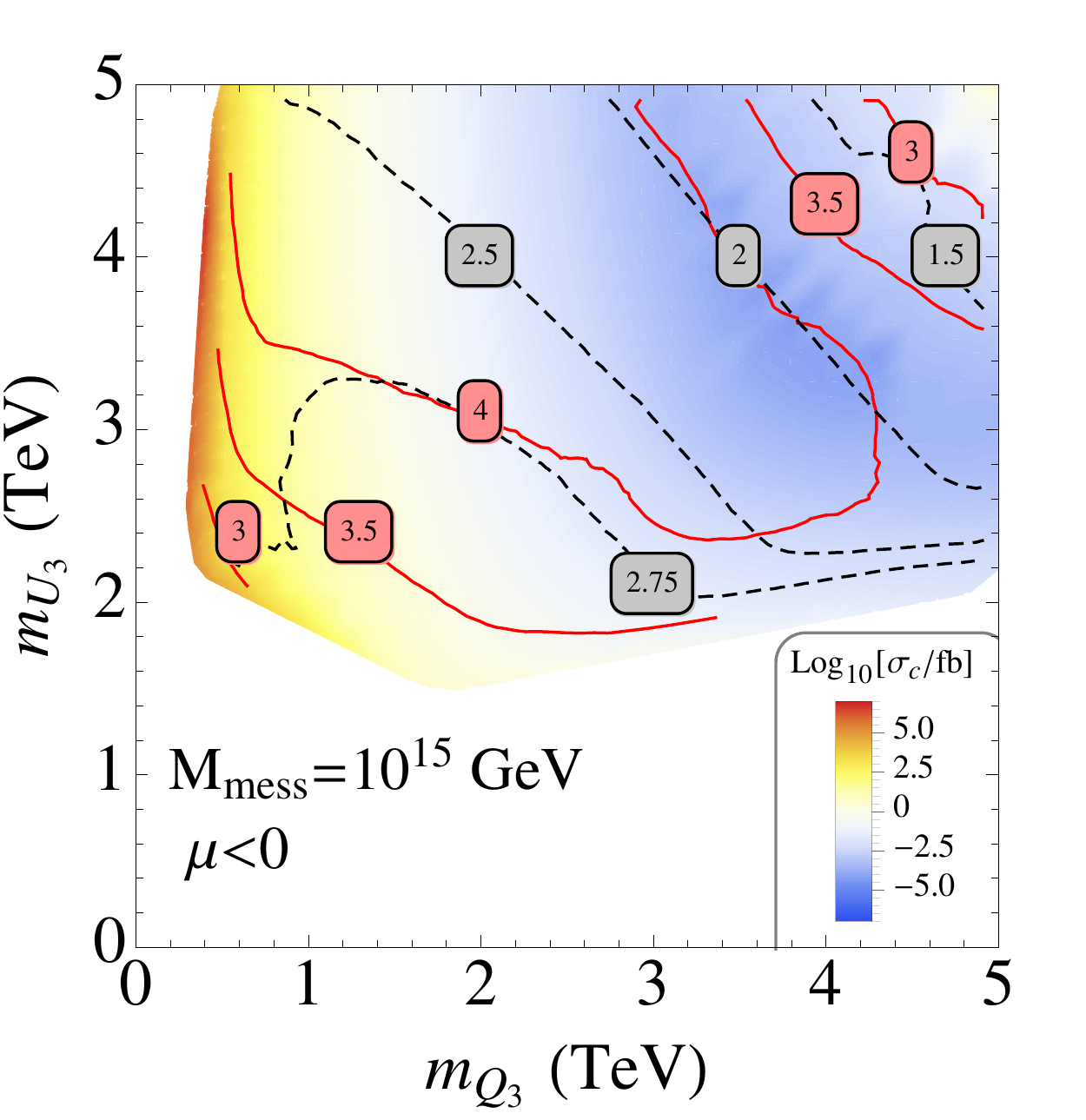}\hfill
\includegraphics[width=0.42\textwidth]{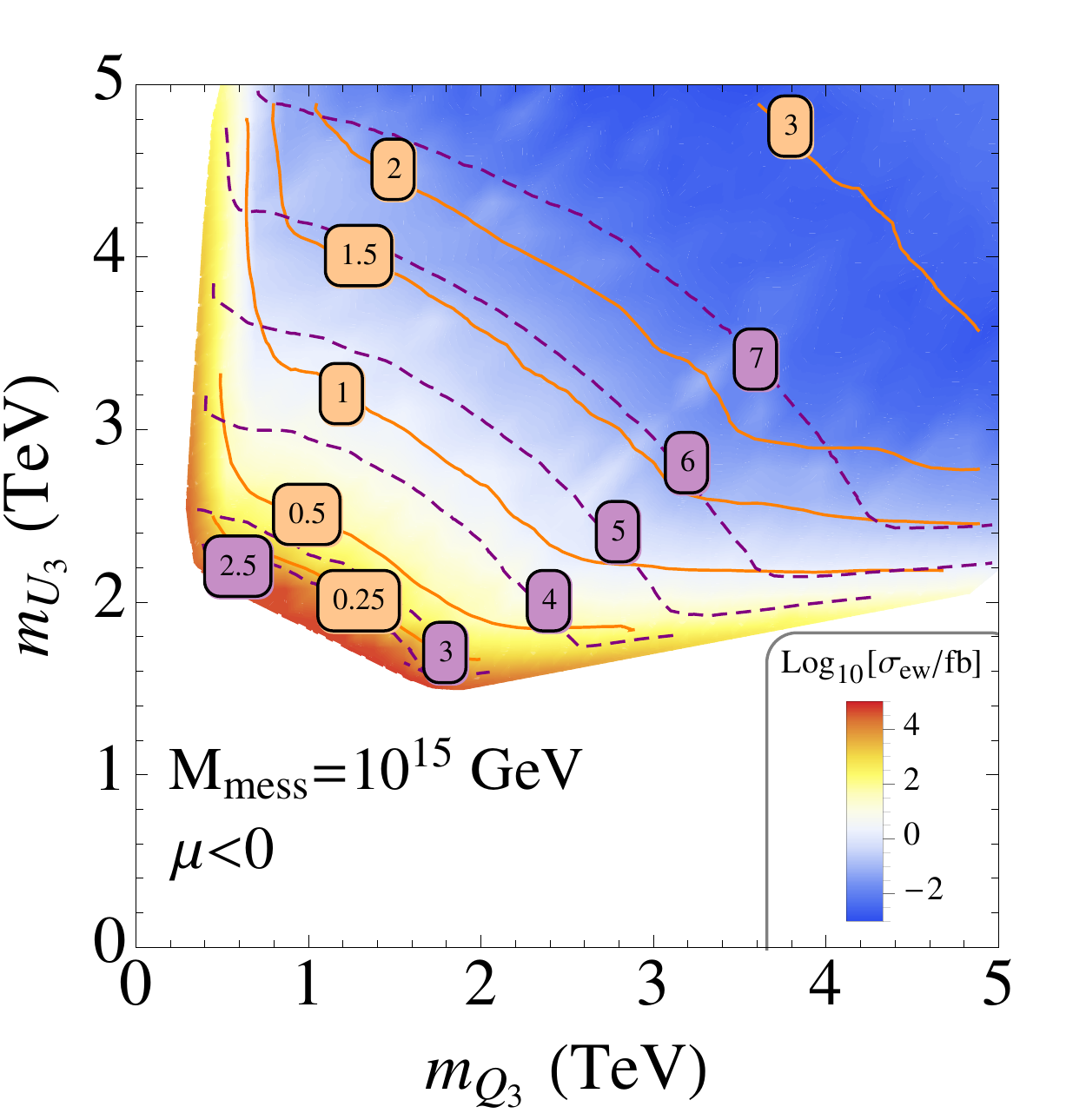}
\includegraphics[width=0.42\textwidth]{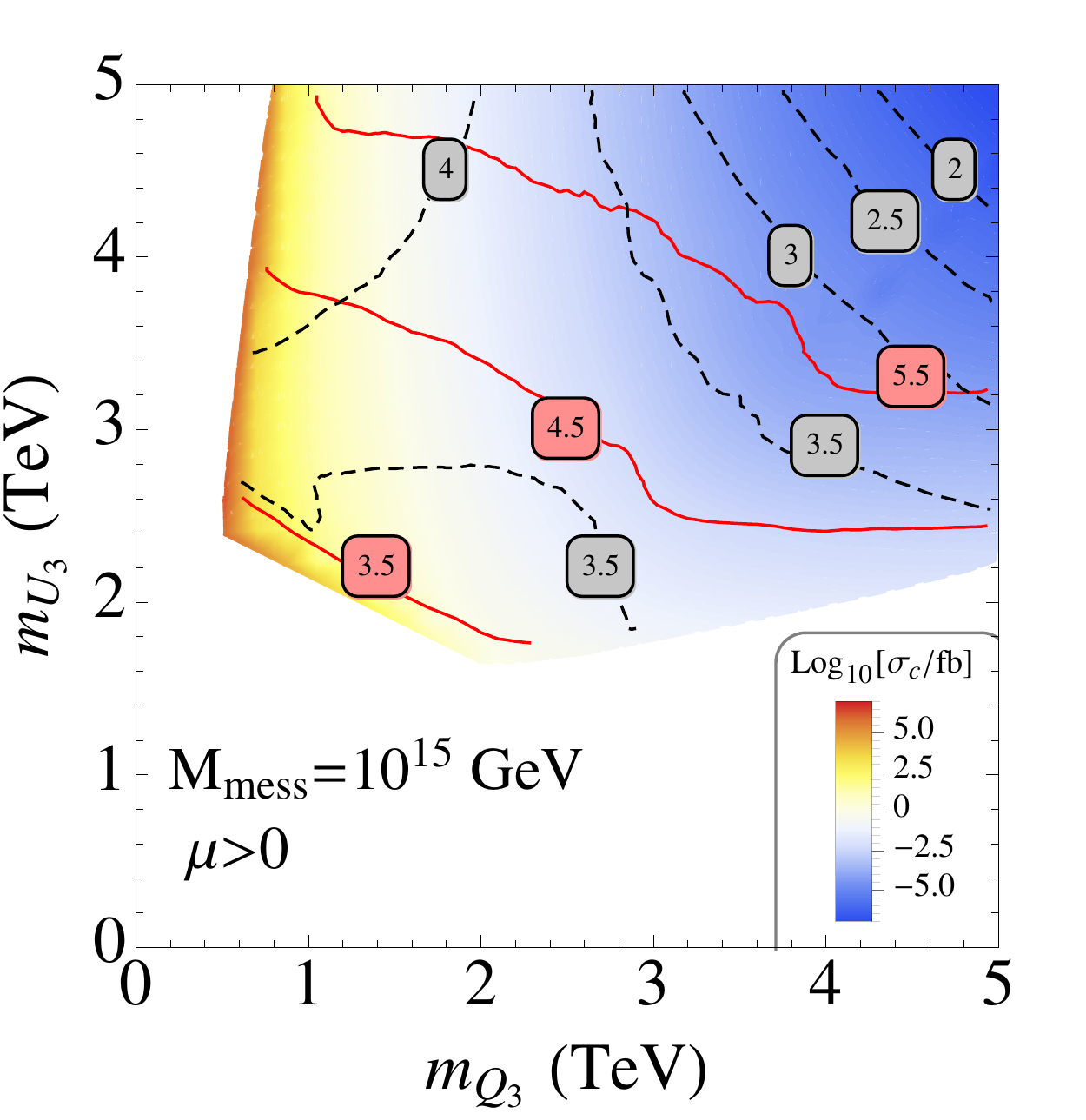}\hfill
\includegraphics[width=0.42\textwidth]{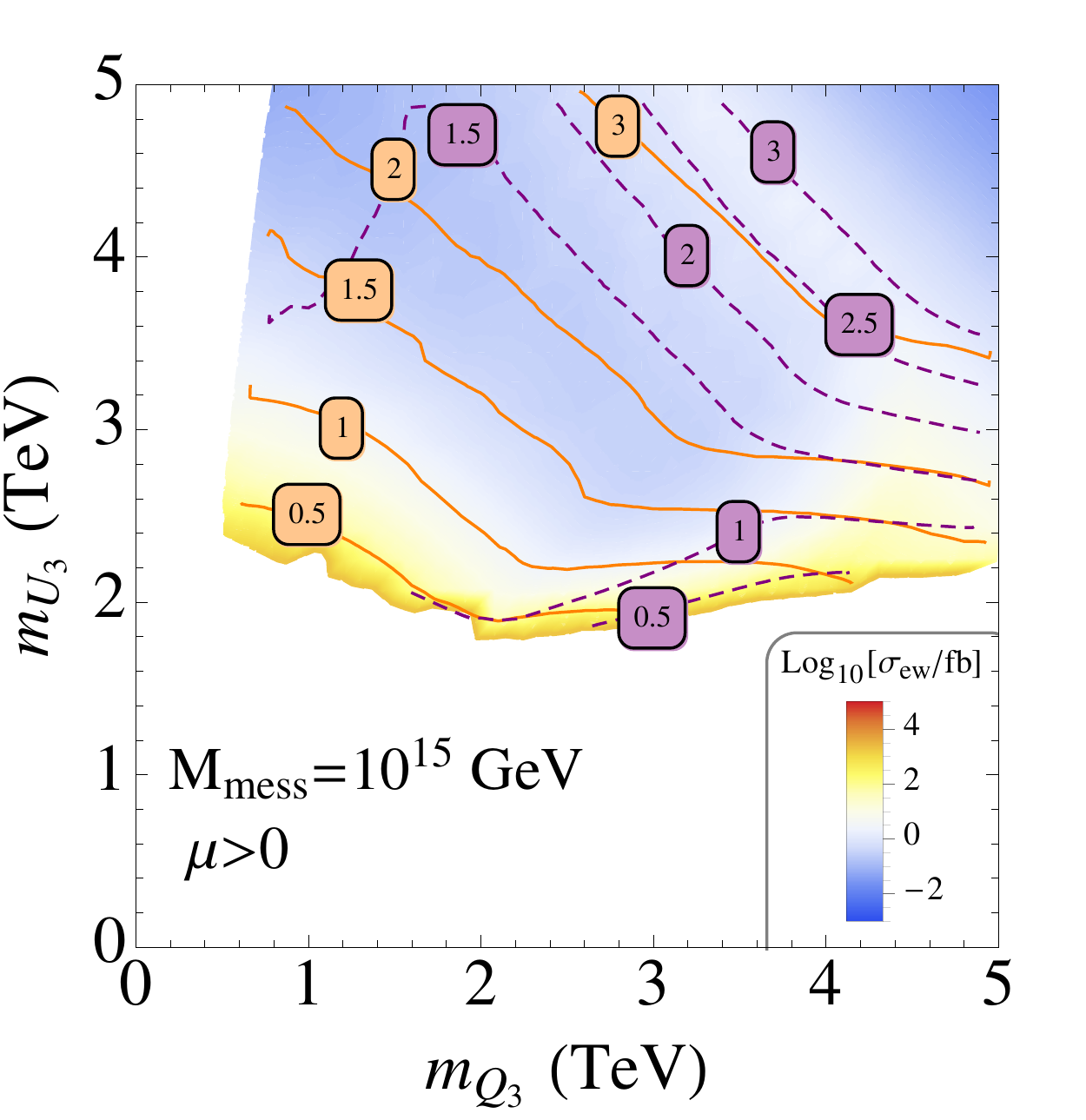}
\includegraphics[width=0.42\textwidth]{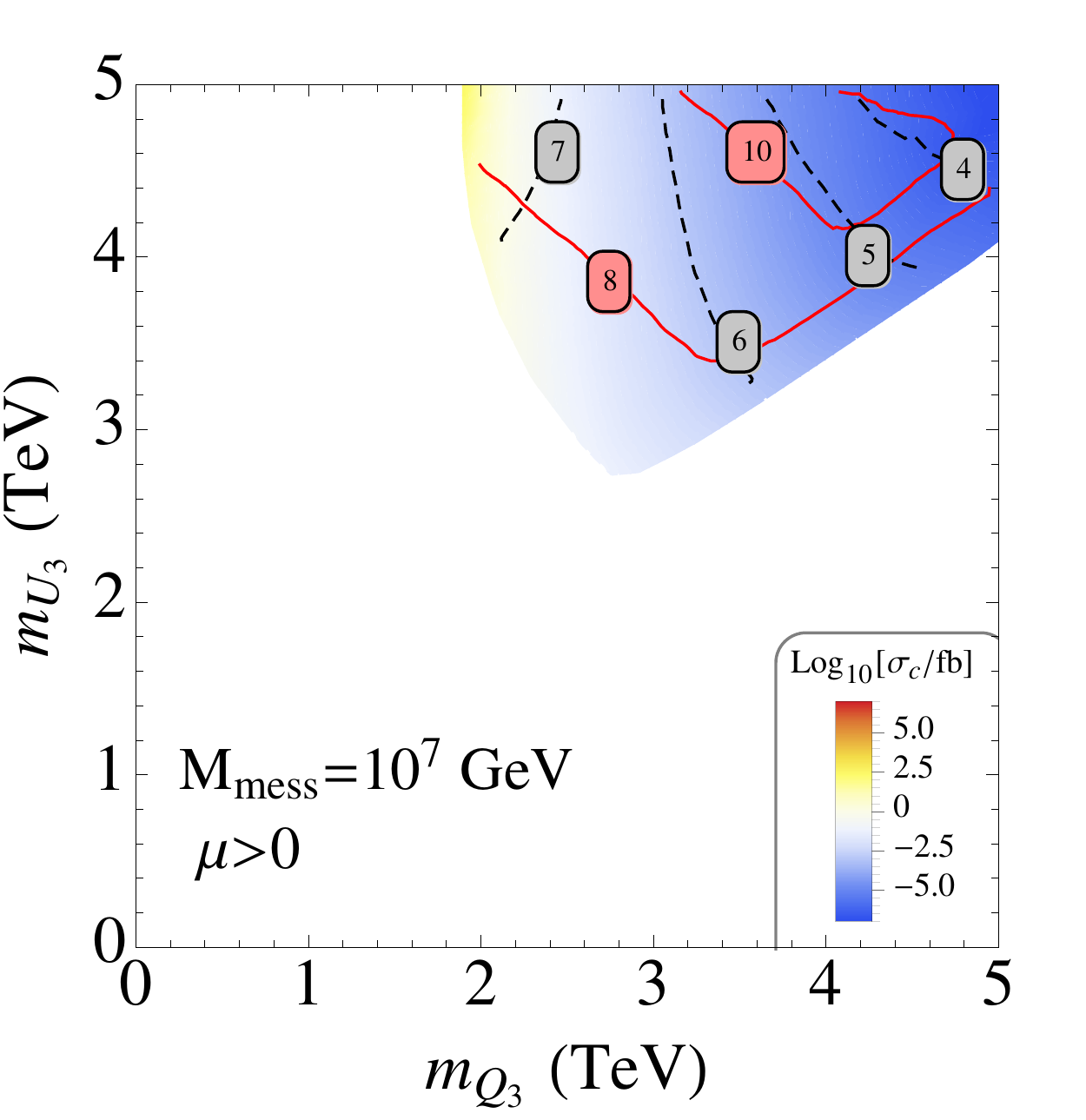}\hfill
\includegraphics[width=0.42\textwidth]{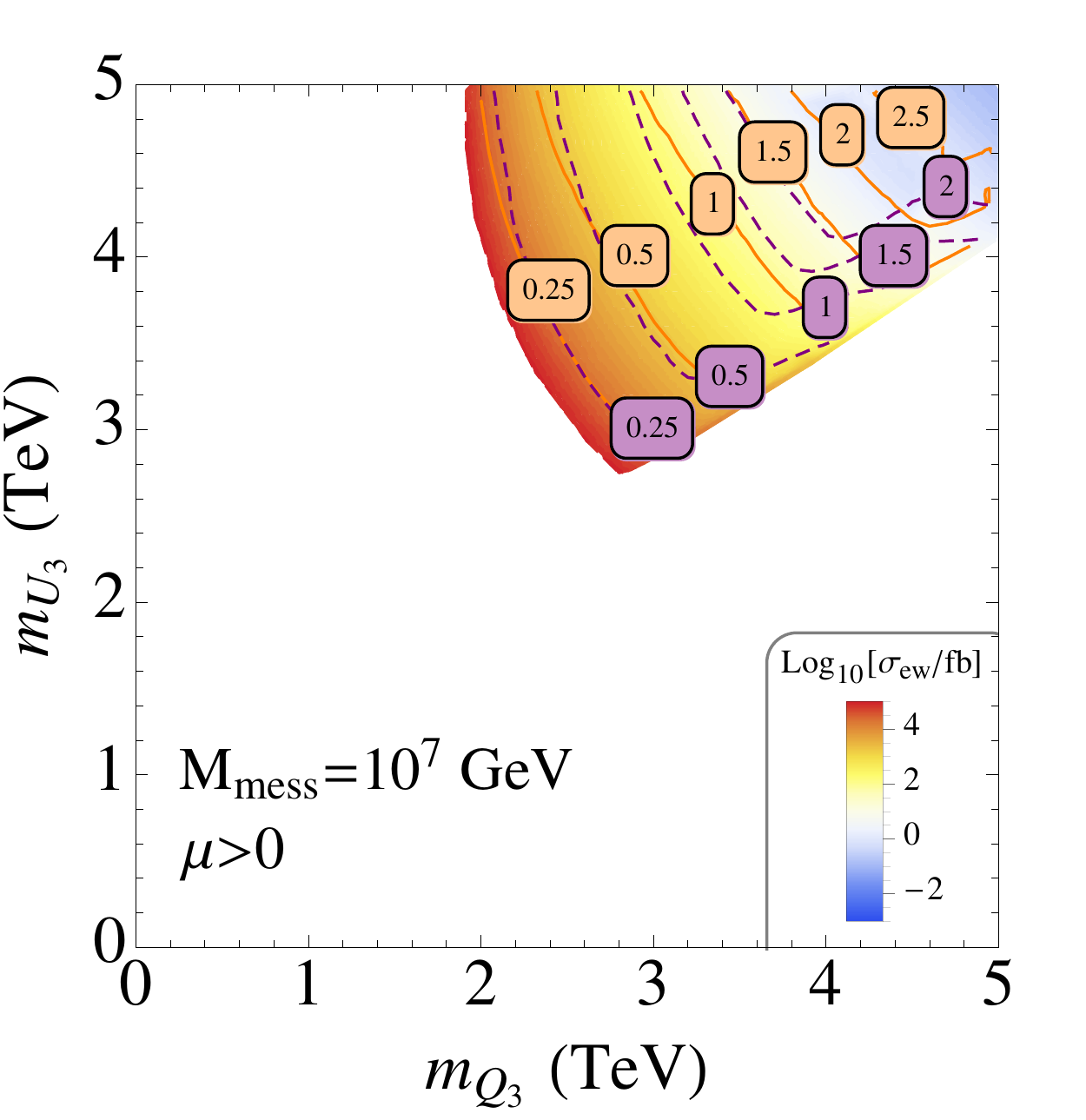}
\caption{In the left/right column: contours of the maximal and minimal $m_{\tilde{g}}$  (red and black dashed) in \text{TeV} and of the maximal $|\mu|$ and $|M_2|$ in \text{TeV} (solid orange and purple dashed). The color shading on the left/right column is the minimal colored/electroweak cross section for $\sqrt{s}=$14 TeV. The $m_{\tilde \chi_1^\pm}\gtrsim 90$ GeV bound from LEP is imposed in the dataset. \label{hwplots}}
\end{figure}

In addition to these simple, $M_{mess}$ independent relations, there are a number of features which do depend on the choice of $M_{mess}$ in a fundamental way. Ultimately, all these features can be traced back to the Higgs mass constraint: For stop masses in the few TeV range, the Higgs in the MSSM can only be sufficiently heavy if the top $A$-term ($A_t$) is large. Since GGM predicts $A_t=0$ at the messenger scale, the $A$-term must be generated in the RG-running, which requires a heavy gluino and/or high messenger scale \cite{Draper:2011aa}.  In \cite{Knapen:2015qba} we showed how this arrangement has the following indirect effects: 
 \begin{itemize}
 
 \item The RG-equation for $m_{H_u}^2$ depends strongly on both $A_t$ and the stop masses, where the net result is that the radiative corrections tend to drive $m_{H_u}^2$ upwards in the regime of interest. In combination with the GGM sum-rule in \eqref{GGMsumrules}, this produces a tension between proper EWSB and the absence of slepton tachyons. In practice this results in a strong lower bound on the stop masses, as shown in fig.~\ref{moneyplane}, where we projected the full parameter space on the plane of the IR stop  soft masses ($m_{Q_3}$ vs $m_{U_3}$). The bound is asymmetric in the stop mass plane due to the $m_E$ relation in \eqref{IRexact}.

 \item Since the horizontal boundary in fig.~\ref{moneyplane} is due to the conflict between achieving EWSB and avoiding a slepton tachyon, both the Higgsino and the sleptons are constrained to be light nearby the boundary.\footnote{We slightly oversimplified the discussion here, since the limits $\mu\to0$ and $m_L\to0$ are in reality not fully independent. We refer to \cite{Knapen:2015qba} for a careful discussion of the correlations induced by sub-leading corrections in $1/\tan\beta$.} For $m_{Q_3}<m_{U_3}$($m_{Q_3}>m_{U_3}$) the lightest slepton is left(right)-handed, again due to \eqref{IRexact}. The sleptons, being scalars, decouple relatively quickly, however the Higgsino must remain light in a substantial part of the parameter space, as shown in fig.~\ref{hwplots}.

\item The milder lower bound on $m_{Q_3}$ has a different origin. Since the gluino must be heavy to ensure a sufficiently large $A_t$, as shown in fig.~\ref{hwplots}, the small tree-level stop masses receive a large gluino-induced threshold correction \cite{Pierce:1996zz}, which tends to drive the stop tachyonic. In this part of parameter space there is effectively a little hierarchy problem between the stop and the gluino. This effect is especially pronounced for $M_{mess}=10^7$ GeV, since the gluino must be much heavier in this case. 

\item There is an interesting correlation between the sign of $\mu$ and the gaugino spectrum, which greatly affects the phenomenology.  In particular, we find that $M_2$ is bounded from above (below) for $\mu>0$ $(\mu<0)$. (These bounds can be traced back to a combination of the Higgs mass constraint, the EWSB conditions and the requirement that $B_\mu$ vanishes at the messenger scale.) This implies that the wino can only be light if $\mu>0$, as shown in fig~\ref{hwplots}. Moreover the one loop RG-equation for $A_t$ is\begin{equation}\label{AtLLog}
\frac{d A_t}{d t}= y_t\left(\frac{32}{3}g_3^2 M_3 + 6 g_2^2 M_2+\frac{26}{15}g_1^2 M_1\right)\ ,
\end{equation}
and recalling that $A_t=0$ at the messenger scale, \eqref{AtLLog} indicates an anti-correlation between $M_2$ and $M_3$ for a fixed $A_t$ at the weak scale. (The $M_1$ contribution is subleading for $M_1$ in the few TeV range.) This anti-correlation implies that the viable range of $M_3$ depends on the sign of $\mu$ as illustrated by the contours of the maximum and minimum gluino mass in fig~\ref{hwplots}: For $\mu<0$ the gluino mass tends to be lighter than for $\mu>0$. 

\item Finally, if both wino and the Higgsino are light, they induce a small but positive one loop contribution to $m_h$ \cite{Pierce:1996zz}. The impact of this correction is especially important for $ M_{mess}=10^7$ GeV, since a large $A$-term is more difficult to achieve in this case. As discussed in the previous bullet point, this spectrum is only possible if $\mu>0$.  As a consequence there is no parameter space for LHC-accessible squarks with $\mu<0$ and $ M_{mess}=10^7$ GeV.
\end{itemize}
 
\subsection{The LEP bound}

Before discussing the different GGM scenarios for the LHC and their corresponding limits, we take a moment to stress the importance of the limits set already by LEP.  Due to the nature of the experiments, searches for SUSY at LHC are necessarily less inclusive and more complex than the corresponding searches at LEP. For the GGM parameter space, the relevant limits are those on the chargino's, as the sleptons are almost always heavier than the kinematic reach of LEP. Even though the LEP limit on the lightest chargino is more inclusive than most LHC searches, there is still some dependence on the full chargino/neutralino spectrum. To account for this, the LEP experiments analysed a wino-like and a Higgsino-like benchmark model and found limits between roughly 90 and 105 GeV \cite{pdg,LEPSUSYgroup}, depending on the model and the $\tilde \chi^\pm_1$-$\tilde \chi^0_1$ mass splitting. Throughout this paper we impose the conservative limit of $m_{\tilde \chi_1^\pm}\gtrsim 90$ GeV. We find that the constraints on the parameter space are not significantly modified by a $\sim 10$ GeV shift in this limit.

Surprisingly, these LEP limits provide a fairly strong constraint on the GGM parameter space, despite their very limited mass reach. This can best be seen from fig.~\ref{moneyplane}, where the LEP limits on the charginos exclude stop masses far outside the reach of LEP, and sometimes even outside the reach of the LHC. The reason lays in the upper bounds GGM enforces on the Higgsino and wino masses, as discussed in the previous section and as depicted in fig.~\ref{hwplots}. This leads to a particularly strong constraint for $M_{mess}=10^7\text{ GeV}$, as in this case both the wino and Higgsino must be light in much of the parameter space, to accommodate the Higgs mass constraint. For a thorough proof of this effect we again refer to \cite{Knapen:2015qba}.

\subsection{SUSY cross section}

The constraints on the GGM spectrum summarized in the discussion above have a strong impact on the dominant production channels in GGM and consequently on the asymptotic reach of HL-LHC. In fig~\ref{hwplots} we show the minimal cross section at the 14 TeV LHC, separately for colored and electroweak production (indicated as $\sigma_c$ and $\sigma_{ew}$ in the plot), where we marginalized over the full parameter space. We computed the NLO cross section for colored sparticles with prospino2 \cite{Beenakker:1996ch}, while  for the electroweakinos we compute LO cross section with pythia 8  \cite{Sjostrand:2006za,Sjostrand:2007gs}. Since the gluino mass is above 2.5 TeV in most of the parameter space, the colored cross section is predominantly controlled by the squarks and is mostly independent of the sign of $\mu$ and $M_{mess}$. However comparing the two signs of $\mu$ for $M_{mess}=10^{15}$ GeV, we see that the colored cross section is somewhat larger for $\mu<0$ because of the lighter gluino mass with respect to $\mu>0$.
The strong correlation between the squark masses of different families encoded in (\ref{IRexact}) implies that $m_{Q_3}$ and $m_{U_3}$ provide a good parametrization of the colored cross section. This is clearly borne out in the left-hand column of fig.~\ref{hwplots}. For $M_{mess}=10^{15}$ GeV, the lower bound on $m_{U_3}$ implies that the cross section is largest for small $m_{Q_3}$, where it is dominated by the left-handed squarks $\tilde q_L$, $\tilde t_L$ and $\tilde b_L$. For $M_{mess}=10^7$ GeV this continues to be true, however the strong lower bounds on both $m_{Q_3}$ and $m_{U_3}$ result in a very suppressed cross section for colored production after the LEP bounds are imposed. As expected, the minimal cross section decreases as the squarks become heavier. For $\mu<0$ and $m_{Q_3}\sim m_{U_3}\gtrsim 4.5\text{ TeV}$ the minimal colored cross section however increases again, since gluino mass is bounded from above by the Higgs mass constraint: In this part of the parameter space, a larger gluino mass would lead to a large enough $A$-term to \emph{overshoot} the Higgs mass. To some extent this is an artifact of our parametrization where we fixed $\tan\beta=20$, and this effect can be compensated by lowering $\tan\beta$ as the stops become heavier. In this case the minimal colored cross section would continue to decrease as expected. 
 
In contrast with the colored cross section, the electroweak cross section does depend strongly on the sign of $\mu$ and $M_{mess}$ (left-hand column of fig.~\ref{hwplots}). For $\mu<0$ the wino is decoupled and the minimal electroweak cross section tracks the maximal Higgsino mass (for example for $\mu\sim 1\text{ TeV}$ we see that $\sigma_{ew}\sim 1\text{ fb}$). More interestingly, for $\mu>0$ the wino and the Higgsino can both be light and both contribute to the electroweak cross section.  This plays a particularly important role for the collider phenomenology of $M_{mess}=10^7$ GeV, where colored production is suppressed.  

\subsection{The NLSP in GGM}
After giving an overview of the dominant production channels in the stop mass plane, it is time to isolate the characteristic decay topologies of GGM. This task is generally challenging since searches at the LHC depend strongly on multiple features of the spectrum. A natural organizing principle in gauge mediation is given by the fact that the gravitino is always the LSP, such that the nature of the NLSP is of particular importance for the collider phenomenology.

While the NLSP is guaranteed to decay to the gravitino plus a standard model state, the time scale of this decay may vary over many orders of magnitude. In particular, the NLSP decay tends to occur outside the detector unless the messenger scale is low (i.e. $M_{mess}\lesssim10^7$ GeV), in which case the decay may occur prompt, displaced or outside the detector. In this paper we focus on the phenomenology of long-lived NLSPs: Once the Higgs mass is imposed, this is the most generic scenario, under some  fairly broad conditions on the nature of the SUSY-breaking sector (see  appendix \ref{app:NLSPdecay}).

We assume R-parity conservation, such that the NLSP is produced in pairs, either through direct production or through a cascade decay. The associated signatures strongly depend on the SM quantum numbers of the NLSP itself, specifically whether the NLSP is charged under the strong and/or electromagnetic force:
  \begin{itemize}
 \item A \emph{neutral NLSP} (neutralino or sneutrino) escapes the detector without leaving any track and is only seen as missing transverse energy. Its direct production at a hadron collider can generically only be bounded by mono-jet searches \cite{Khachatryan:2014rra,Aad:2015zva}. (Although sometimes more efficient search strategies can be designed if the NSLP is a member of a quasi-degenerate electroweak multiplet, see below.) The strongest bounds therefore almost always come from cascade decays of other, heavier superpartners. 
 \item On the other hand, a \emph{charged and/or colored NLSP} (stop, sbottom, stau or gluino) produces a spectacular signature in the form of a pair of highly ionizing tracks, and searches for this final state are nearly background free \cite{Chatrchyan:2013oca,ATLAS:2014fka}. Extra activity in the event from a cascade is therefore not needed to further reduce the background, although a cascade decay from a colored state may still greatly increase the cross section in the case of a stau NLSP.
 \end{itemize}
\begin{figure}\centering
\includegraphics[width=0.45\textwidth]{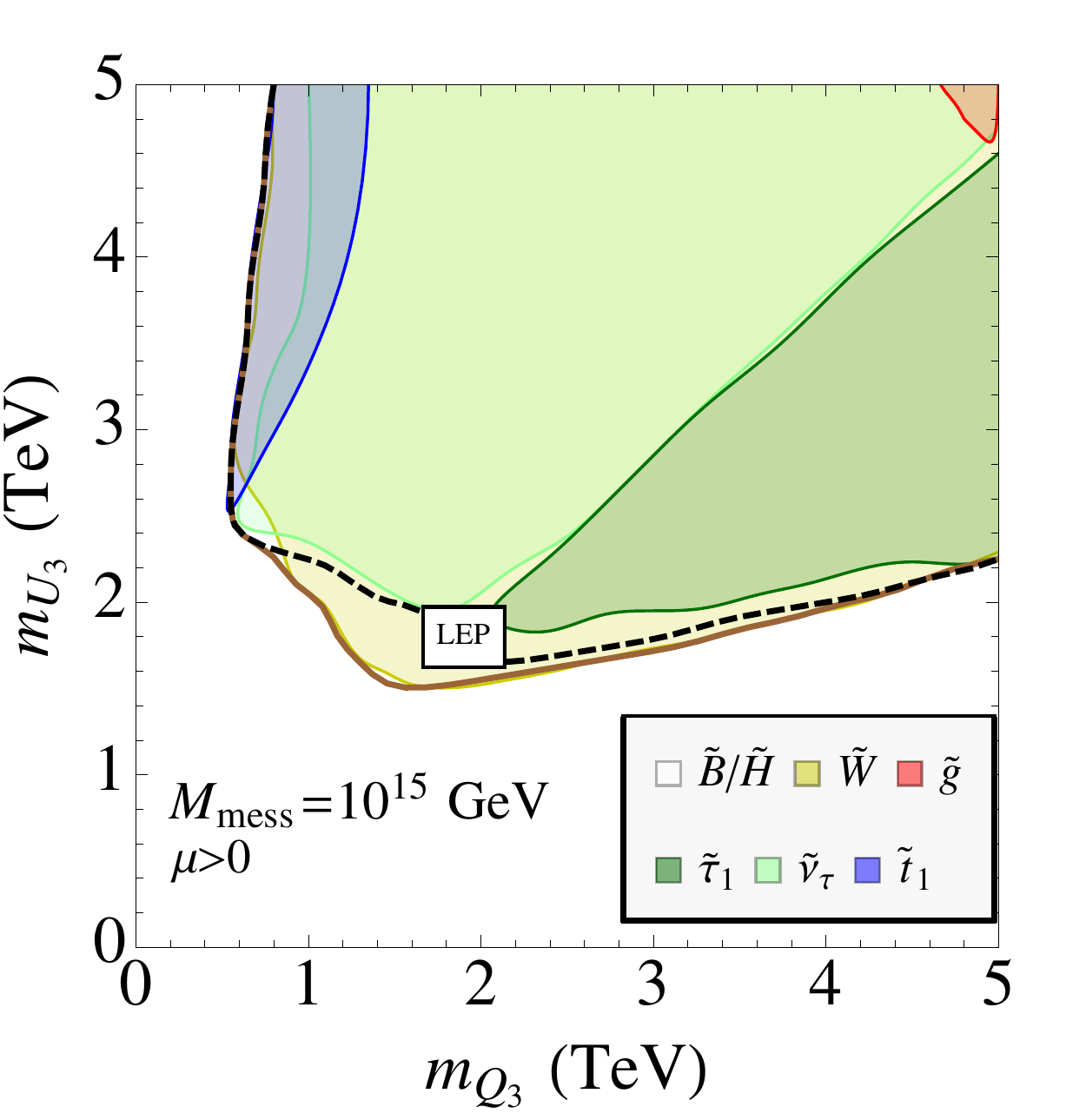}\hfill
\includegraphics[width=0.45\textwidth]{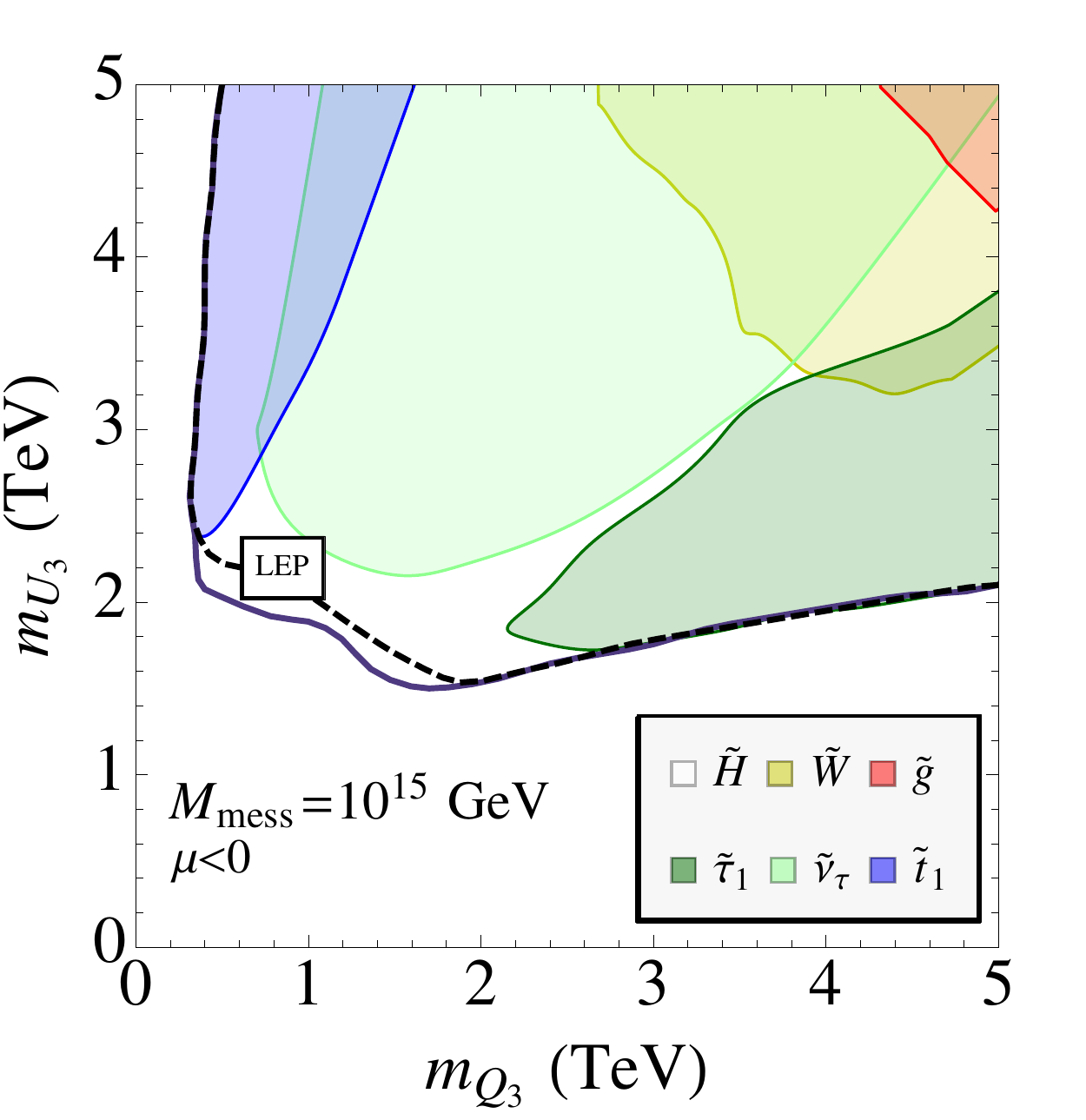}\hfill
\includegraphics[width=0.45\textwidth]{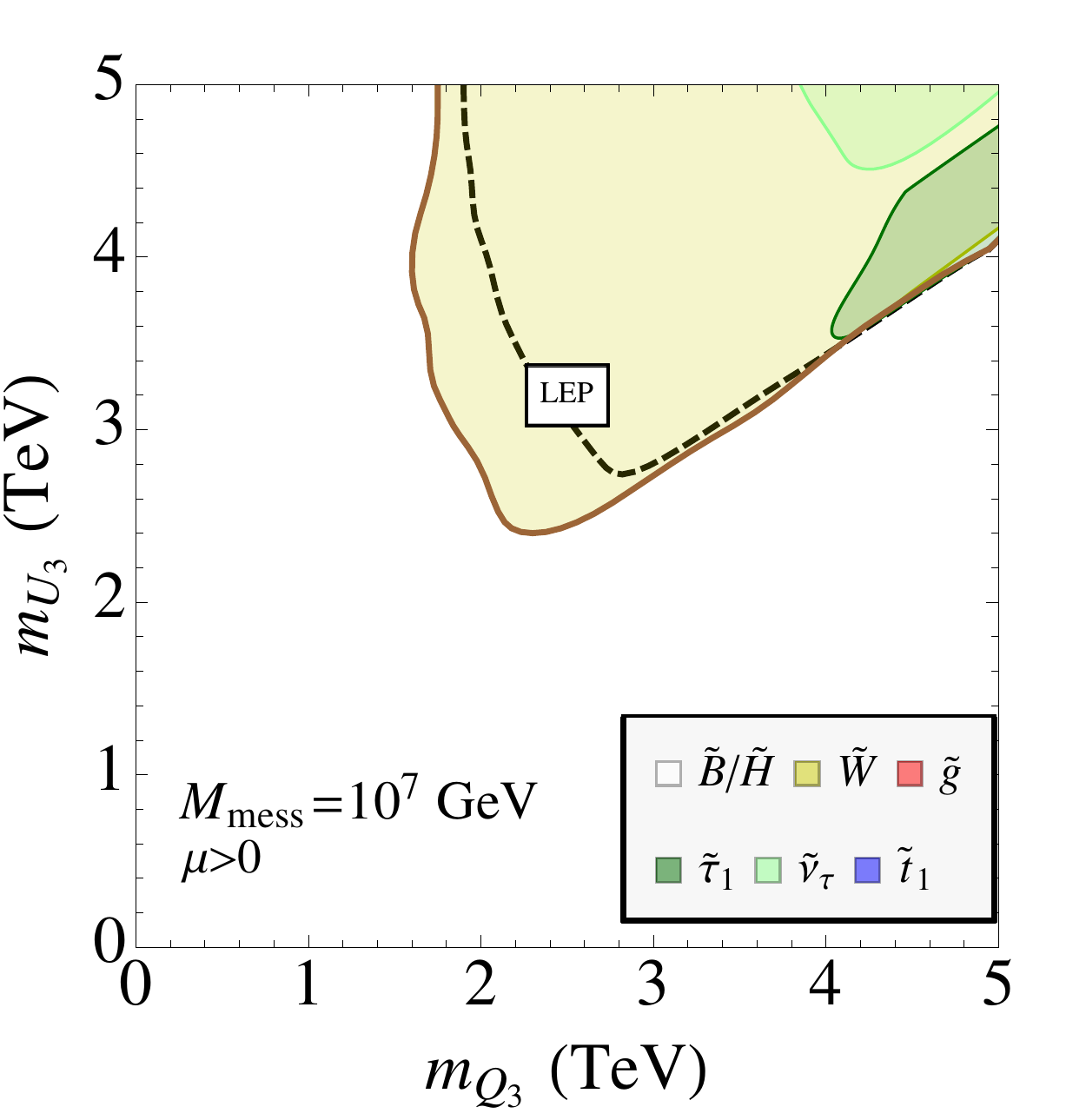}
\caption{Allowed NLSP types in the stop soft mass plane. The bino and Higgsino NSLP are possible everywhere within the allowed region and are not explicitly plotted.  \label{NLSPplots} For $M_{mess}=10^{15}$ GeV, the NLSP tends to be long-lived, while for $M_{mess}=10^{7}$ GeV it may decay prompt, displaced or outside the detector. The precise lifetime is model dependent, see appendix \ref{app:NLSPdecay}. }
\end{figure}

In fig.~\ref{NLSPplots} we show how requiring a particular NLSP type correlates with the stop masses. This correlation, together with the total SUSY cross section shown in fig.~\ref{hwplots}, already provides a rough idea of the reach of LHC at 14 TeV for a given NLSP. Before discussing this in more detail, we stress that the projection of the possible NLSP types on the $m_{Q_3}$ vs $m_{U_3}$ plane in fig.~\ref{NLSPplots} only shows whether a given NLSP type is allowed for certain values of the squark masses, which roughly correspond to the colored cross section (see fig.~\ref{hwplots}). Fig.~\ref{NLSPplots} should therefore not be interpreted as a measure of the relative abundance of a given NLSP in the full parameter space. Since certain UV completions may very well favor one NLSP type over another, we will treat every allowed NLSP in GGM on the same footing in this paper.  

The \textbf{neutralino NLSP} is possible for any value of the stop masses. In general the lightest neutralino is a mixture between the neutral components of the bino, wino and Higgsino, but in most of the parameter space these masses are relatively split, such that the NLSP can usually be thought of as either mostly bino, wino or Higgsino. Depending on the relative hierarchy among $M_1$, $M_2$ and $\mu$ a neutralino NLSP can be:
\begin{itemize}
\item  \textbf{Bino-like:}  if $|M_1|<|\mu|,|M_2|$. The bino NLSP plays a special role in our discussion, as $M_1$ only enters in the RG-equations only through terms suppressed by $\sim g_1^2$. As a result, $M_1$ is an approximate flat direction in the parameter space which does not significantly affect the rest of the spectrum as long as $M_1$ is in the TeV-range. Since there is no direct limit on a bino NLSP (not even from LEP if $\mu$ is heavy enough \cite{Calibbi:2013poa}) a bino NLSP is allowed everywhere in the parameter space.  
Note however that near the lower bound on $m_{U_3}$,  $\mu$ and $M_2$ (if $\mu>0$) are bounded from above, such that NLSP is necessarily somewhat mixed, even if $|M_1|<|\mu|,|M_2|$.

\item \textbf{Higgsino-like} if $|\mu|<|M_{1,2}|$. A Higgsino-like NLSP is possible for all values of the stop masses. Its mass is bounded from above as was shown in fig.~\ref{hwplots}, and bounded from below by the LEP bound of $m_{\tilde C_1}>103.5$ GeV. The Higgsino multiplet consist of two neutral states and one charged state which are typically split between 1 and 10 GeV for $M_1$ in the TeV range. The charged state decays promptly to the lightest neutral state but the standard model particles in this decay are generally too soft to benefit from dedicated search strategies at LHC \cite{Gori:2013ala,Baer:2014cua}. 

\item \textbf{Wino-like} if $|M_2|<|\mu|,|M_1|$. As was mentioned earlier and shown in fig.~\ref{hwplots}, for $\mu<0$ $|M_2|$ is bounded from below and the wino can only be the NLSP if the stop masses are both above 3 TeV and $M_2$ is larger than 2 TeV. Conversely, for $\mu>0$ the wino may be the NLSP in nearly all of the stop mass plane. Also in this case there is a strong upper bound on $M_2$ for stops close to the boundary of the viable region (see fig.~\ref{hwplots}). This bound is particularly strong for $M_{mess}=10^7$ GeV. The wino multiplet consists of a charged ($\chi_1^\pm$) and a neutral state ($\chi_1^0$), typically split by $\sim$ a few 100 MeV to a few GeV, depending on whether the Higgsino is nearby in the spectrum. $\chi_1^+$ decays to $\chi_1^0$ by emitting a pion, which is too soft to be a useful observable, as was the case for the Higgsino NLSP. An important exception can arise if the splitting is smaller than $\sim$~200~MeV, in which case the decay is displaced and $\chi_1^\pm$ is seen as a disappearing track \cite{CMS:2014gxa,Aad:2013yna}. 
\end{itemize}
For the Higgsino and wino NLSP scenarios we always assume a decoupled bino for simplicity. If the bino mass were instead in the proximity of the Higgsino or wino NLSP mass, the splitting between the components of the multiplet would typically increase.

A {\bf slepton NLSP} can be either mostly left-handed (L) or right-handed (E). The third generation sleptons are typically slightly lighter than the first two generations, such that a slepton NSLP is usually a right-handed stau or a left-handed snutau.\footnote{For certain very small corners of the parameter space it is also possible that the slepton hierarchy is reversed, which results in a selectron or smuon NLSP. We will not discuss this case separately in this paper, and refer the reader to \cite{Calibbi:2014pza} for more details on this scenario.} From the GGM sum-rules (\ref{GGMsumrules}) we thus expect a $\tilde \tau_R$ NLSP for $m_{Q_3} > m_{U_3}$ and $\tilde \nu_\tau$ NLSP for $m_{Q_3} < m_{U_3}$, which is very clearly confirmed by our numerical solution in fig.~ \ref{NLSPplots}. 

If $\tilde \tau_R$ ($m_{Q_3} \gtrsim m_{U_3}$) is the NLSP, the final state consists out of a pair of highly ionizing charged tracks \cite{CMS-PAS-EXO-15-010,ATLAS:2014fka}. In the region where this configuration is possible, the stop and the gluino are however fairly heavy, and the dominant contribution to the cross section comes from electroweak states. In the region where the left-handed slepton is the lightest slepton ($m_{Q_3} \lesssim m_{U_3}$) the NLSP is always the neutral $\tilde \nu_\tau$ since the charged states of the $L$ doublet are always pushed upwards by a small but irreducible D-term contribution of the form
\begin{equation}
m_{\tilde{\ell}}-m_{\tilde{\nu}}\simeq \frac{m^2_{W}\sin^2\beta}{m_{\tilde{\ell}}+m_{\tilde{\nu}}}\ .\label{splittting}
\end{equation} 
The splitting between the slepton and the sneutrino is usually around $10-30$ GeV and it decreases with increasing sneutrino mass. Direct production of slepton pairs will therefore lead to soft leptons in association with missing energy. The snutau NLSP behaves mostly like a neutralino NLSP, and the only bound on its direct production comes from LEP \cite{LEPSUSYgroup}. However, since some fine tuning is involved in order to get the sneutrino light, it is very atypical to get it within the reach of LEP experiments (practically speaking its mass never goes below $\sim 150\text{ GeV}$ in our dataset). Fortunately, for a snutau NLSP the squarks may be light enough to make this scenario testable at LHC. 

A \textbf{stop or sbottom NLSP} does not occur for $M_{mess}=10^{7}$ GeV, barring very fine-tuned regions in the parameter space. They can however be realized for $M_{mess}=10^{15}$ GeV if $m_{Q_3} \ll m_{U_3}$, but even in this case  its mass is still always smaller than roughly 1.5 TeV  (again see \ref{NLSPplots}).  Since the stop/sbottom NLSP must always be left-handed, the second colored state in the $SU(2)$ multiplet is always nearby in mass. 

 Finally, the \textbf{gluino} can always be the NLSP in the (mini-)split regime where the scalars are heavy. (In this case no large $A$-term and therefore no large $M_3$ is needed, since the Higgs mass comes predominantly from the mass of the stops.)  If the gluino is the NLSP, the squarks are never accessible at the LHC. Since this case only arises in a very small part of our parameter space and is studied elsewhere already \cite{Giudice:2004tc,ArkaniHamed:2004fb,ArkaniHamed:2004yi,ArkaniHamed:2012gw,Arvanitaki:2012ps,Cohen:2015lyp}, we do not discuss it in any more detail.

\clearpage

\section{Neutral NLSPs\label{sec:neutralNLSP}}

In this section we present the most relevant simplified topologies for the neutral NLSPs in GGM, and estimate the current constraints and the future reach of the High Luminosity LHC (HL-LHC) with 3000 $\text{fb}^{-1}$. For this purpose we take the neutral NLSPs to be long-lived on detector length scales, which is  preferred for $M_{mess}\gtrsim10^{7}$ GeV under some fairly broad assumptions about the nature of the SUSY-breaking sector (see appendix  \ref{app:NLSPdecay}). For $M_{mess}\lesssim 10^{7}$ GeV, the neutral NLSP may decay either prompt, displaced or outside the detector. A visible NLSP decay results in two additional SM bosons in the final state, which typically strengthens the limits on a neutralino NLSP. (For a discussion of prompt decays of neutralino NLSP's we refer to \cite{Meade:2009qv,Ruderman:2011vv,Kats:2011qh} and to \cite{Meade:2010ji,Liu:2015bma} for an analysis of displaced decays.) On the other hand, a sneutrino NLSP decays fully invisibly to a neutrino and a gravitino, and in this case the collider phenomenology therefore does not depend on the NLSP lifetime.

The collider phenomenology depends on whether the NLSP is a sneutrino or a neutralino with a dominant wino, Higgsino or bino component. Moreover many features of the spectrum drastically change with the choice of  $M_{mess}$ and the $\text{sign}(\mu)$, as we discussed in the previous section. We therefore  organize our discussion around the NLSP type, $M_{mess}$ and sign($\mu$). For each case we discuss the most important simplified topologies and present the mass range for each particle in the SUSY spectrum which is compatible with the Higgs mass constraint, the GGM boundary conditions, proper EWSB and LEP constraints. We subsequently overlay our estimates for the constraints from the existing LHC data, as well as an estimate of the future sensitivity of the HL-LHC. 

It is always possible to trivially relax all collider constraints by simply decoupling the whole superpartner spectrum, and for concreteness we therefore require $M_s\equiv\sqrt{m_{\tilde t_1}m_{\tilde t_2}}<4$ TeV. This allows us to focus on the ``low'' $M_s$ phenomenology of gauge mediation, where at least some of the scalar superpartners are within the reach of LHC and where the gluino does not sensibly contribute to colored sparticle production. For $M_s>4$ TeV the squarks are decoupled but the gluino is not, and the phenomenology strongly resembles that of (mini)-split supersymmetry  \cite{Giudice:2004tc,ArkaniHamed:2004fb,ArkaniHamed:2004yi,ArkaniHamed:2012gw,Arvanitaki:2012ps,Cohen:2015lyp}. The most important collider bounds for long-lived neutral NLSPs come from SUSY cascades initiated by colored or electroweak superpartners. In our treatment of the collider bounds we neglect the efficiency drop of SUSY searches for squeezed spectra (i.e. when the NLSP mass is nearby the mass of the sparticle responsible for the production). Since we effectively marginalize over the NLSP mass, including squeezing would imply that the limits would be effectively dominated by the squeezed spectra. Such spectra only occur in a small part of the parameter space, and this would therefore result in an overly pessimistic picture of the constraints on the parameter space as a whole. For completeness, we give an example of how much squeezed spectra can affect the constraint on the GGM parameter space with Higgsino NLSP in appendix \ref{app:squeezed}.

\subsection{Wino NLSP}

\begin{figure}[t]\centering
\includegraphics[width=\textwidth]{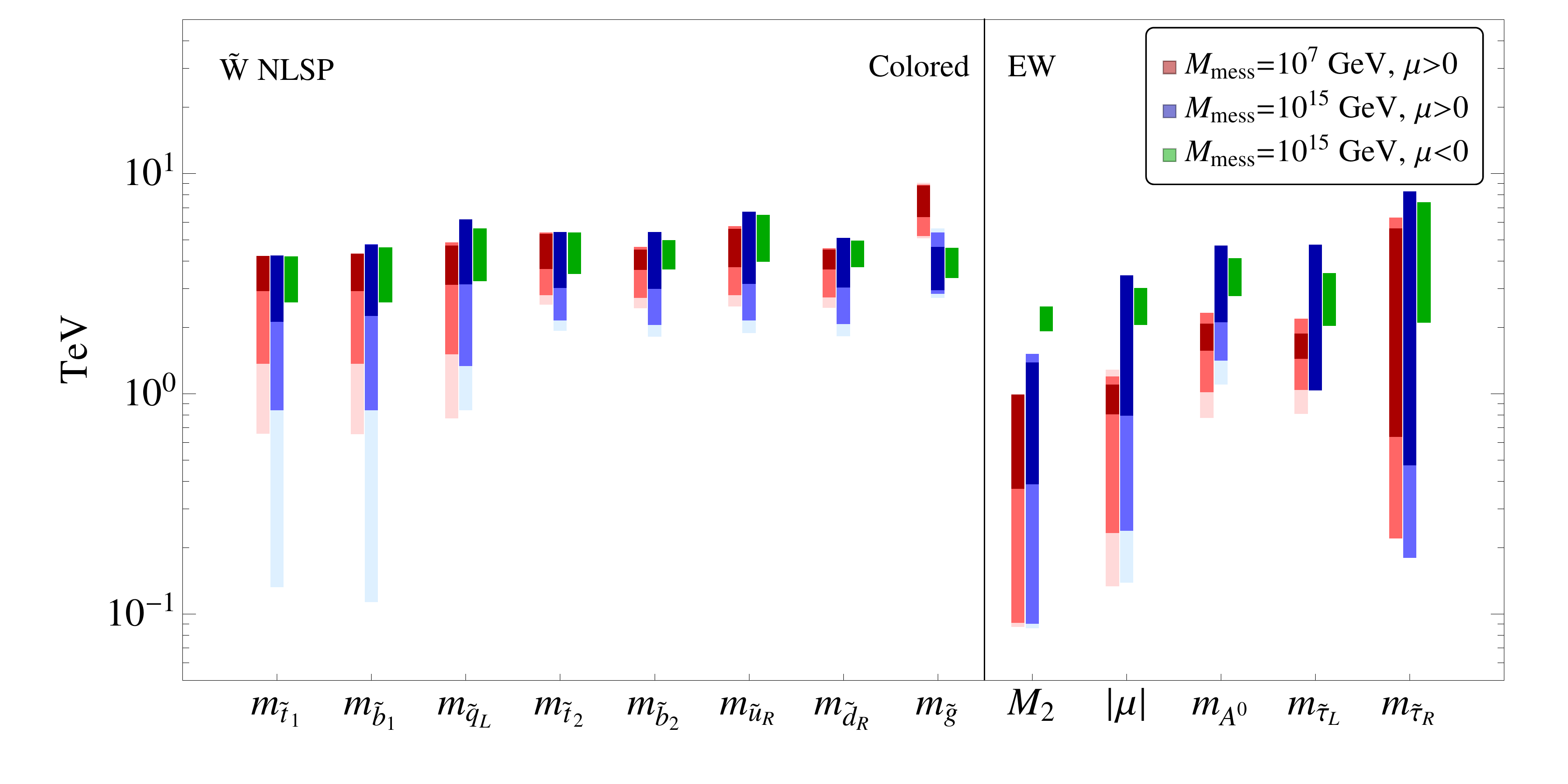}
\caption{Overview of the GGM parameter space with $M_s<4$ TeV for a wino NLSP. Regions with lightest shading are excluded or disfavored by existing data, regions with darker shading are accessible at the HL-LHC, while darkest regions are likely to be unaccessible at the LHC. See main text for details.}  \label{villadoroW}
\end{figure}

The structure of the GGM spectrum with a wino NLSP is summarized in fig.~\ref{villadoroW}, where the various bands indicate the viable interval for each mass parameter. The most important feature is that a wino NLSP within the LHC reach is only possible if $\mu>0$. (See also fig.~\ref{hwplots} and fig.~\ref{NLSPplots}.) In the remainder of this section we therefore only focus on $\mu>0$.

Having fixed $M_1$ to be decoupled, the splitting between charged ($\tilde\chi^\pm_1$) and neutral ($\tilde\chi^0_1$) components of the wino multiplet depends on the Higgsino mass and may be sufficiently small for the disappearing track searches to become relevant \cite{CMS:2014gxa,Aad:2013yna}. If the splitting between $\tilde \chi^+_1$ and $\tilde \chi^0_1$ is too large for the decay to be displaced, the decay product is typically too soft to provide a good enough handle to separate the SUSY signal from the SM background. Both $\tilde \chi^+_1$ and $\tilde \chi^0_1$ can then effectively be treated as MET. In this case the bounds on the wino NLSP scenario thus come exclusively from cascade decays, as we will detail in the next subsection. 

After the collider bounds are imposed, fig.~\ref{villadoroW} reveals two particularly interesting correlations in the spectrum. Firstly, for HL-LHC there are very mild \emph{upper} bounds on the wino and gluino masses. This is a result of the anti-correlation between $M_2$ and $M_3$, discussed in sec.~\ref{sec:solveGGM}, which indicates that a lower bound on $M_2$ implies an upper bound on $M_3$ and vica versa. A second interesting feature is that the left-handed slepton and $A^0$ are both forced to be heavy, which holds regardless of the collider constraints. This is again a consequence of the Higgs mass constraint and the EWSB conditions \cite{Knapen:2015qba}. The LHC further constrains these particles indirectly through the limit on the wino mass. Finally, in \cite{Knapen:2015qba} it was shown that for $\mu>0$, $\mu$ increases if $M_2$  increases. This correlation results in a strong, indirect lower bound on $M_2$ from the LHC bound on $\mu$ (see next subsection).

In summary, we see that once future HL-LHC bounds are accounted for, the only states below one TeV are electroweakinos, controlled by $\mu$, $M_2$ and the right-handed sleptons. 

\subsubsection{Constraints and simplified topologies}

We first discuss the bounds on direct wino production. The disappearing track search is relevant if $m_{\tilde C_1}-m_{\tilde N_1}\lesssim$ 200 MeV  \cite{CMS:2014gxa,Aad:2013yna}, which corresponds to $\mu\gtrsim500\text{ GeV}$. The electroweak production of one or two disappearing tracks is therefore always dominated by direct wino production. The bound on disappearing tracks from the 8 TeV (HL-LHC) data on a directly produced wino ranges between $100$ GeV (300 GeV) for $m_{\tilde C_1}-m_{\tilde N_1}\sim 200$ MeV up to $500$ GeV (1300 GeV) for $m_{\tilde C_1}-m_{\tilde N_1}\sim 140$ MeV.   For the HL-LHC reach we obtained a naive bound by rescaling the projected bound obtained in \cite{Low:2014cba,Cirelli:2014dsa} to different values of $m_{\tilde C_1}-m_{\tilde N_1}$ by assuming the same expected sensitivity as in \cite{CMS:2014gxa}.  When imposing the bound on disappearing tracks we are neglecting a possible enhancement of the signal strength coming from left-handed squarks decaying to the lightest chargino either directly or through a SUSY cascade involving the Higgsino. Recasting and optimizing disappearing tracks searches for SUSY spectra with a non-negligible colored production would not qualitatively modify our conclusions regarding the currently excluded GGM spectra in fig. \ref{villadoroW} but it could provide one of the most sensitive probes of the wino NLSP scenario at the HL-LHC.

 If the splitting $m_{\tilde C_1}-m_{\tilde N_1}\gtrsim$ 200 MeV, the only bound on direct wino production comes from mono-jet and VBF+MET searches, which are very weak even at the HL-LHC \cite{Low:2014cba,Han:2013usa,Berlin:2015aba}. The sparticles higher up in spectrum thus dominate the collider limits, and it is therefore most convenient to think about the phenomenology in terms of simplified topologies. Both the gluino and the right-handed squarks are decoupled as far as the current LHC data is concerned. Therefore the most promising production channels are the Higgsino, the left-handed squarks and the left-handed stop/sbottom. The sleptons tend to be relatively heavy, except very near the boundary of the viable region. Given that their production cross section is also very small, their direct production typically does not play a role in the constraints. 

\begin{figure}[t]\centering
\begin{subfigure}[b]{0.25\textwidth}\centering
\includegraphics[height=3cm]{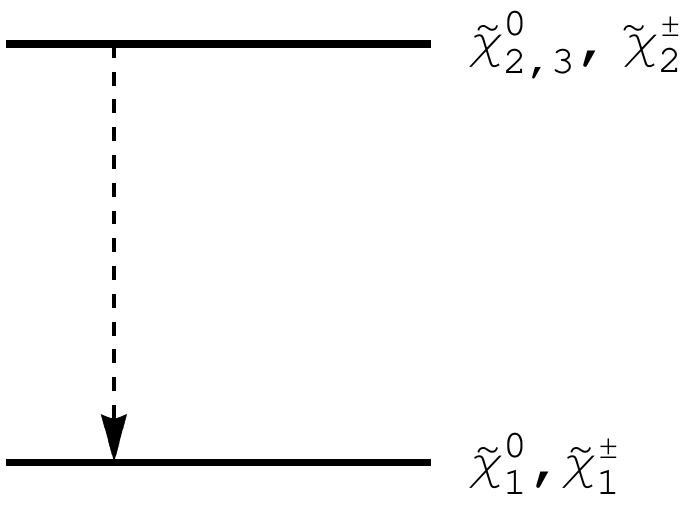}
\caption{Higgsino-wino}  \label{HWtopology}
\end{subfigure}\hfill
\begin{subfigure}[b]{0.23\textwidth}\centering
\includegraphics[height=3cm]{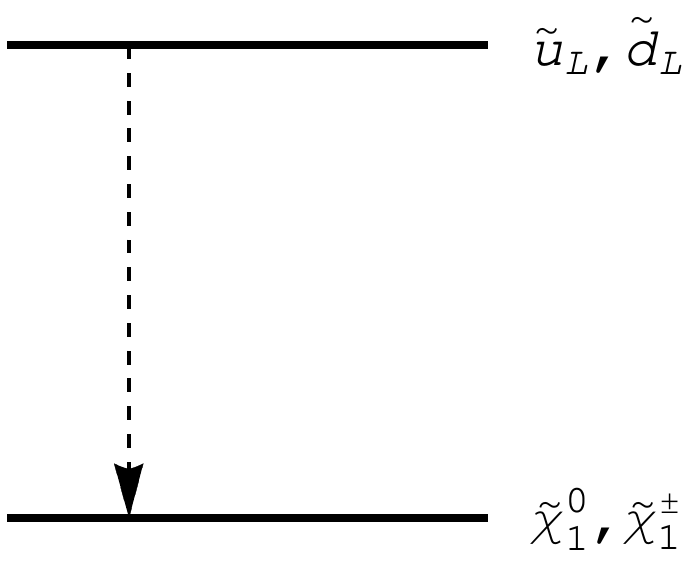}
\caption{$\tilde q_L$-wino}  \label{qLWtopology}
\end{subfigure}\hfill
\begin{subfigure}[b]{0.23\textwidth}\centering
\includegraphics[height=3cm]{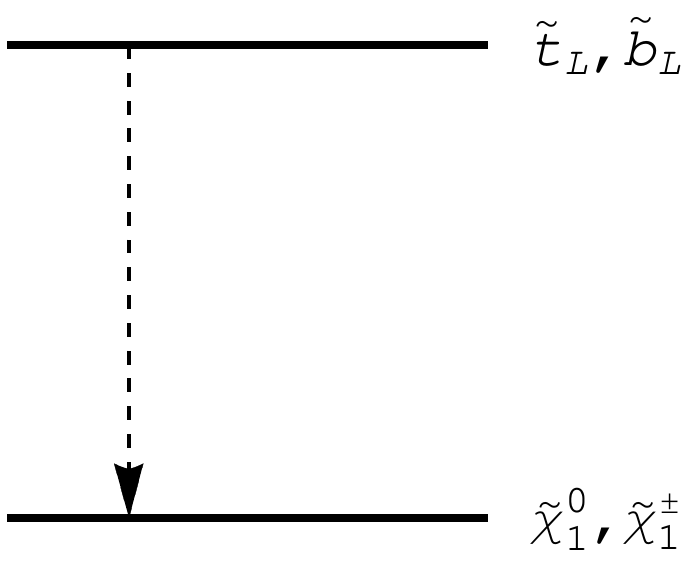}
\caption{ $\tilde t_L/\tilde b_L$-wino }  \label{TWtopology}
\end{subfigure}\hfill
\begin{subfigure}[b]{0.25\textwidth}\centering
\includegraphics[height=3cm]{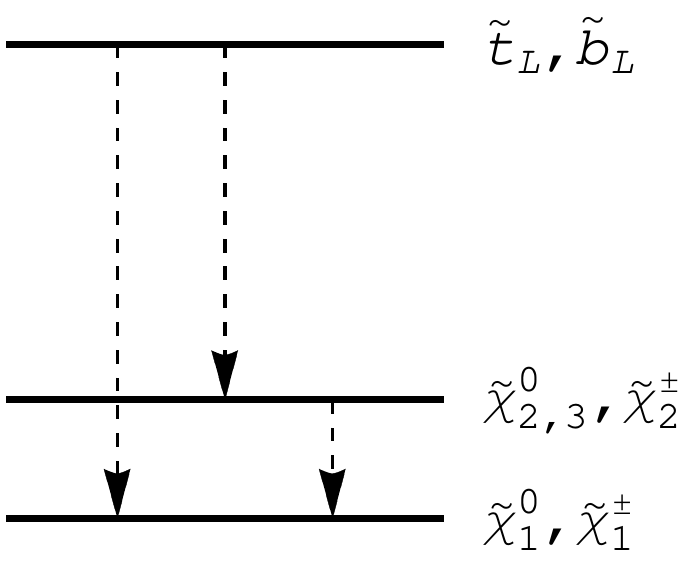}
\caption{$\tilde t_L/\tilde b_L$-Higgsino-wino}  \label{THWtopology}
\end{subfigure}
\caption{Most relevant simplified topologies for the wino NLSP.}\label{fig:Winotopo}
\end{figure}

\begin{figure}[t]
\includegraphics{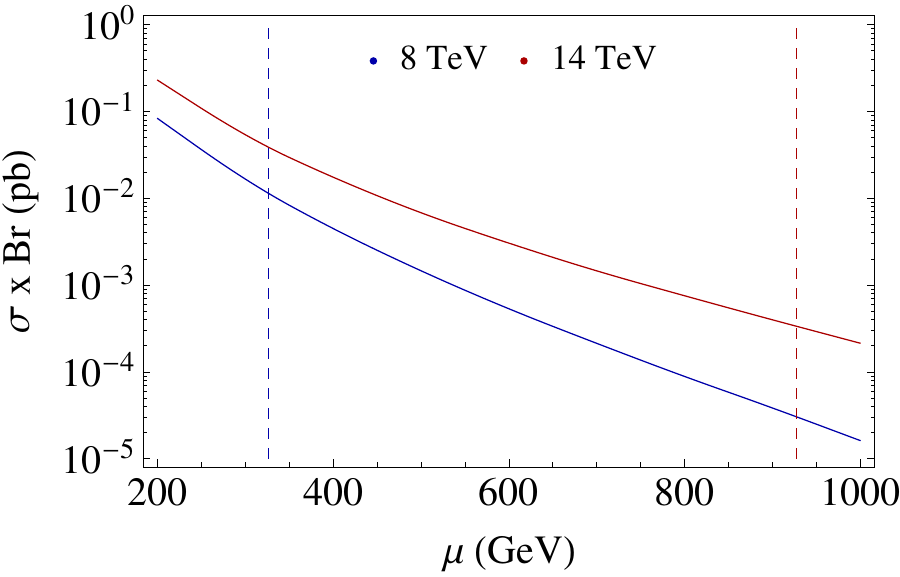}
\caption{Effective rate for the $WZ$+MET final state from Higgsino production with a wino NLSP, with $M_2=100$ GeV. The dashed blue (red) lines indicate the current (projected) limits.\label{fig:WZplot}}
\end{figure}

The four most relevant simplified topologies are shown in fig.~\ref{fig:Winotopo}: In the first topology all colored states are decoupled, such that the production cross section is provided by the Higgsino (fig.~\ref{HWtopology}). The decay to the wino NLSP occurs through the emission of gauge and Higgs bosons, of which $WZ$+MET final state is the most constraining. In fig.~\ref{fig:WZplot} we show the effective rate for this process, where we combined all production channels, weighted by the branching ratio to $WZ$+MET. Also indicated on fig.~\ref{fig:WZplot} is the current limit, as well as the projected limit for the HL-LHC, which we rescaled from \cite{Aad:2014vma}  and \cite{ATL-PHYS-PUB-2014-010,CMS-PAS-SUS-14-012} respectively. Throughout this paper, the various branching ratios were computed with SUSY-HIT \cite{Djouadi:2006bz}. 

The second simplified topology (fig.~\ref{qLWtopology}) parametrizes the case where the left-handed squarks are accessible. The branching ratio to jets+MET is 100\%, but we rescale the bound in \cite{Khachatryan:2016xvy} to account for the fact that only the $\tilde q_L$ are accessible.\footnote{In principle there are a number of important caveats to this approach which should be mentioned \cite{Mahbubani:2012qq}. First, the efficiencies of the searches drop dramatically for squark masses below $\sim 700$ GeV. For the right-handed squarks this region is already excluded by the constraints in fig.~\ref{moneyplane}, while for the left-handed squarks this region is excluded by constraints on the left-handed stop and/or sbottom. Secondly, the difference in the parton luminosity functions implies that the cross sections of the $\tilde u$ and $\tilde d$ are significantly higher than that for $\tilde c$ and $\tilde s$. However since the squark spectra in gauge mediation are flavor degenerate in the first two generations, this effect is unimportant.} In this rescaling we made use of the NLO squark production cross section as computed by prospino2 \cite{Beenakker:1996ch}, where we account for the (mild) dependence of the squark cross section on the gluino mass (see fig.~\ref{hwplots}). For gluino masses in the middle of the allowed range, this roughly amounts to $m_{\tilde q_L}\gtrsim1100$ GeV ($m_{\tilde q_L}\gtrsim 1050$ GeV) for $M_{mess}=10^{15}$ GeV ($M_{mess}=10^{7}$ GeV). At the HL-LHC, the $\tilde u_R$ and $\tilde d_R$ may be accessible as well, although they are most likely not degenerate with $\tilde q_L$. In this case we estimate the limits by requiring that the total squark cross section is smaller than the projected limit in \cite{ATL-PHYS-PUB-2014-010}. A priori there is a second squark-initiated topology, where the Higgsino is in between the squarks and the NLSP. However because of the small coupling of the lowest generation squarks to the Higgsino, the branching ratio of the cascade decay via the Higgsino is negligible, and we do not need to consider this case separately. 

\begin{figure}[t]
\includegraphics[width=0.7\textwidth]{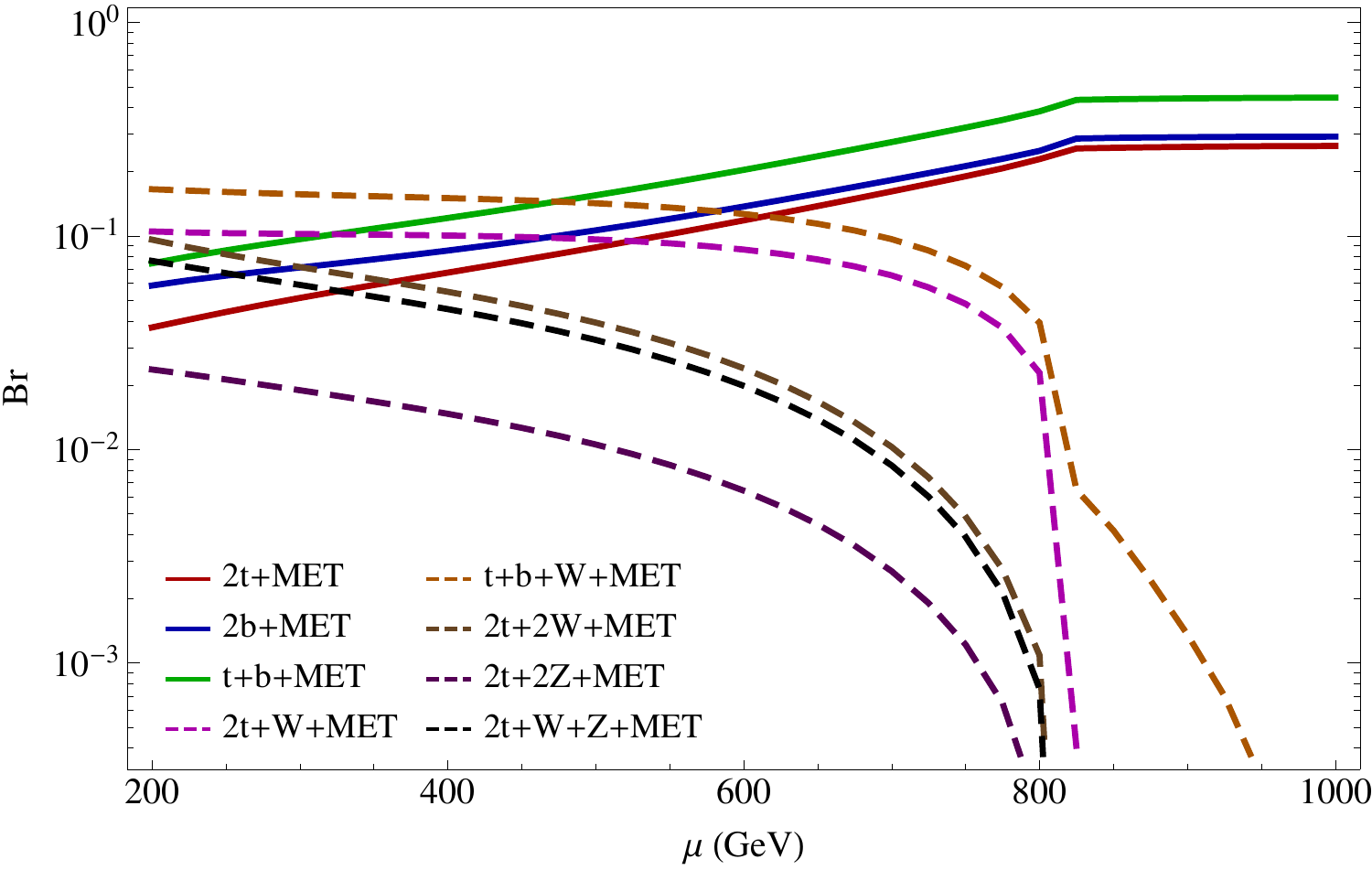}
\caption{Branching ratios for some of the channels in the $\tilde t_L/\tilde b_L$-Higgsino-wino topology, with $m_{\tilde t_L}=m_{\tilde b_L}=1$ TeV and $M_2=100$ GeV.   \label{fig:THWcascade}}
\end{figure}

The final production mode is through the left-handed stop and sbottom. In this case the presence of the Higgsino in the middle of the spectrum does make a qualitative difference. We first consider the case where the Higgsino decouples, as in fig.~\ref{TWtopology}. Even though the $A$-term is large, the stop mixing is still relatively small in the region where the lightest stop and sbottom are accessible with the current data. The reason is again the lower bound on the right-handed squarks, which enforces a sizable mass splitting between the stop gauge eigenstates. The left-handed stop and sbottom are therefore degenerate to good approximation, and as such their production cross sections are approximately equal. 
To establish a bound, we can therefore treat them as a single state with branching ratios which are the average of those of the stop and the sbottom. The branching ratios to do not depend strongly on the mass parameters and are roughly $\mathrm{Br}(bb+\mathrm{MET})\approx0.31$, $\mathrm{Br}(tb +\mathrm{MET})\approx0.44$ and $\mathrm{Br}(tt +\mathrm{MET})\approx0.25$. The strongest bound comes from the sbottom search and set $m_{\tilde t_L,\tilde b_L}\gtrsim850$ GeV \cite{ATLAS-CONF-2015-066,CMS-PAS-SUS-16-001}, where we again assume no appreciable squeezing between $\tilde t_L$($\tilde b_L$) and the NLSP.  The corresponding estimate for the HL-LHC is $m_{\tilde t_L,\tilde b_L}\gtrsim1400$ GeV \cite{ATL-PHYS-PUB-2014-010}. 
The Higgsino can also be in between the wino and stop/sbottom (fig.~\ref{fig:THWcascade}), in which case the situation is considerably more complicated, as cascade decays now occur frequently. A very large variety of final states is possible for this topology. We show the branching ratios  of some important channels in fig.~\ref{fig:THWcascade}, but we do not attempt a proper recasting and combination in this paper. Instead we simply  stick with  $m_{\tilde t_L,\tilde b_L}\gtrsim850$ GeV and $m_{\tilde t_L,\tilde b_L}\gtrsim 1400$ GeV as a crude estimates of the current and projected limits respectively.

With the current data, the right-handed squarks are only accessible in a small part of the parameter space, however this is no longer true with the HL-LHC data set. At this point a set of analogous simplified topologies with the right-handed squarks will be come relevant, in addition to those discussed in this section.

\subsection{Higgsino NLSP}

\begin{figure}[b!]\centering
\includegraphics[width=\textwidth]{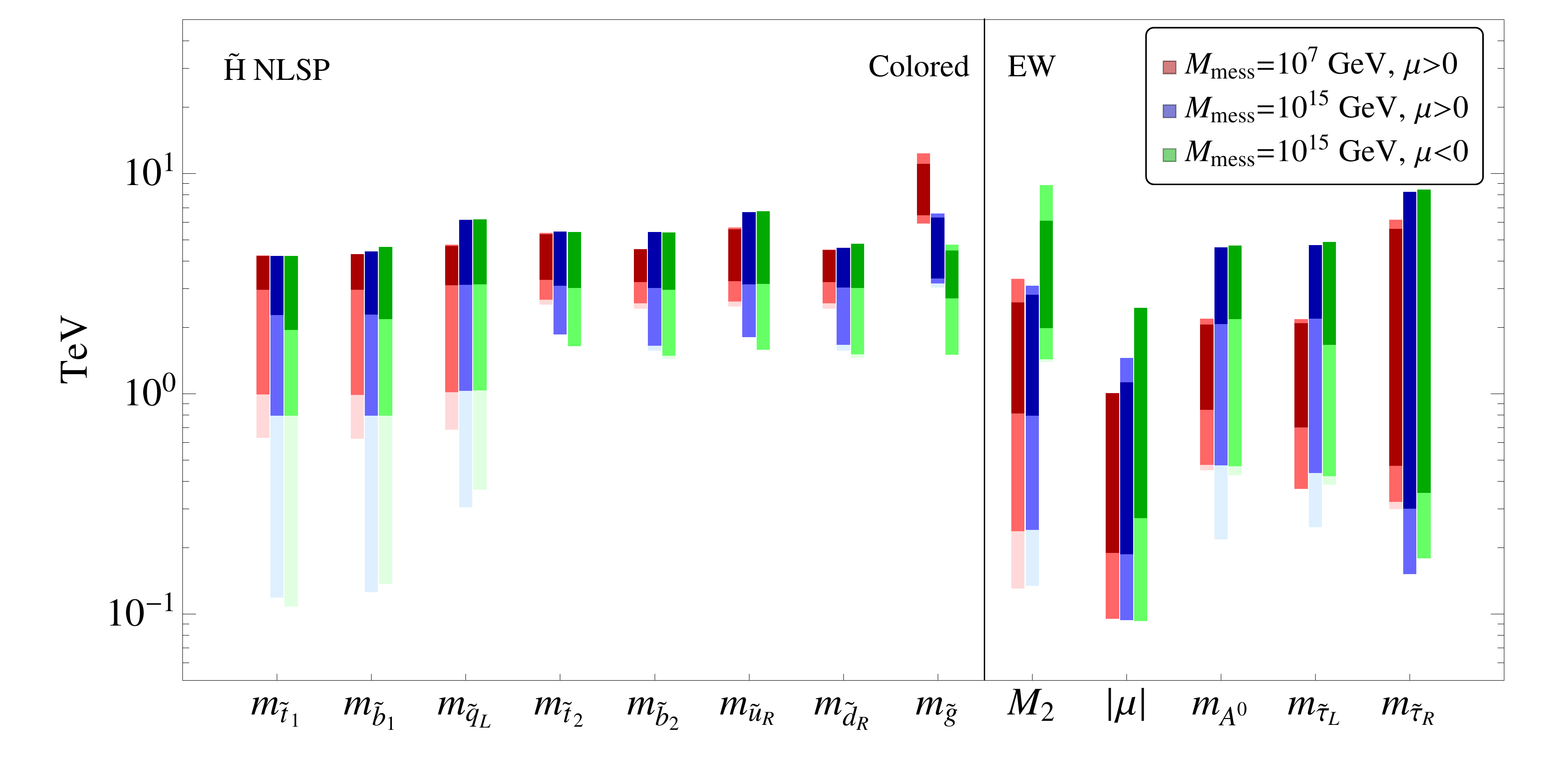}
\caption{Overview of the GGM parameter space for a Higgsino NLSP. Regions with lightest shading are excluded or disfavored by existing data, regions with darker shading are accessible at the HL-LHC, while darkest regions are likely to be unaccessible at the LHC.   }  \label{villadoroH}
\end{figure}

For the Higgsino NLSP, summarized in fig.~\ref{villadoroH}, the gluino and the right-handed squarks are also too heavy to be significantly constrained by the current data.  However in contrast with the wino NLSP, it is possible to probe the $\mu<0$ branch  with current data.   On the other hand, the splitting between the components of the Higgsino multiplet is always large enough to allow for prompt decays of the two heaviest states of the multiplet. The visible decay products of this decay are generally still too soft to be efficiently discriminated from background, such that the strongest limits come from cascade decays of heavier sparticles. We discuss the most relevant topologies in the next section.

As for the wino NLSP, there are a number of GGM-specific correlations in the spectrum. Firstly, the HL-LHC can also set an interesting indirect lower bound on the Higgsino mass of $\sim 300\text{ GeV}$ for the case of $\mu<0$, as is apparent from the summary plot in fig.~\ref{villadoroH}. This bound is sensiblity stronger than the direct bound from monojet+MET \cite{Low:2014cba,Han:2013usa}, which remains applicable for $\mu>0$. The indirect bound is a consequence of the bound on the gluino and a particular correlation in the GGM parameter space which is unique to $\mu<0$ \cite{Knapen:2015qba}. (It is also worth noting that only for $\mu<0$ the HL-LHC projection of the reach on gluino production ($m_{\tilde g}\gtrsim2900$ GeV \cite{ATL-PHYS-PUB-2014-010}) plays a role in constraining GGM.) Secondly, fig.~\ref{villadoroH} reveals the same anti-correlation between $M_2$ and $M_3$ as was present for the wino NLSP. A third example of an important GGM-specific correlation is apparent in the lower bounds on the left-handed sleptons. For $\mu<0$, $M_2$ is always in the multi-TeV range, and keeping the sleptons light therefore requires fine tuning. For $\mu>0$, $m_L$ is correlated with $\mu$ such that the LEP bound on the chargino results in an indirect lower bound on the sleptons. This correlation is strongest for high messenger scales. For more details, we again refer to \cite{Knapen:2015qba}. The lower bound on the left-handed slepton masses subsequently induces a lower bound on $m_{A^0}$ (see eqn.~\eqref{IRexact}).
 
 In summary, we see that the current data do not lead to a strong lower bound on the SUSY spectrum, but the HL-LHC will be able to push every SUSY particle above $1\text { TeV}$ except the Higgsino and the right-handed sleptons, regardless of the choice for $M_{mess}$ and sign$(\mu)$. In addition, for $\mu>0$ the wino can be below 1 TeV. If $M_{mess}$ is also low, the left-handed sleptons and the CP-odd Higgs could be relatively light as well. This feature is most pronounced for the Higgsino and sneutrino NLSP's due to the relation in eqn.~\eqref{IRexact}. The pseudo-scalar and slepton masses also tend to be lower for low messenger scales, which can be understood from combining the RG-running with the EWSB conditions, see eqn.~(2.5) in \cite{Knapen:2015qba}.

\subsubsection{Simplified topologies}

\begin{figure}[t]\centering
\begin{subfigure}[b]{0.3\textwidth}\centering
\includegraphics[height=3.cm]{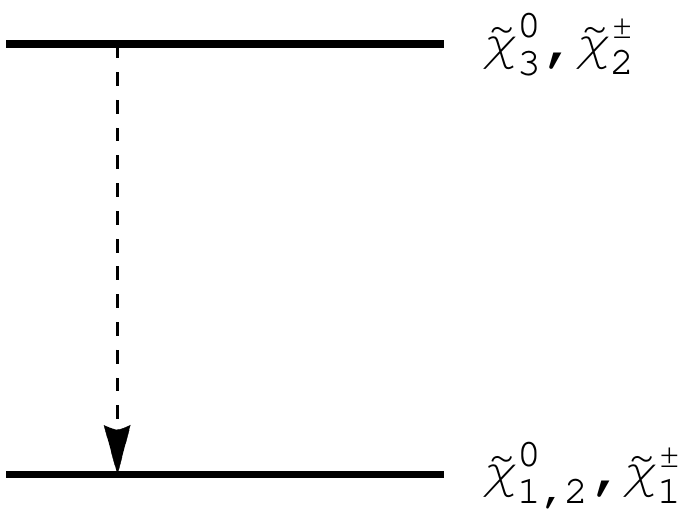}
\caption{ wino-Higgsino}  \label{WHtopology}
\end{subfigure}\hspace{.5cm}
\begin{subfigure}[b]{0.3\textwidth}\centering
\includegraphics[height=3.cm]{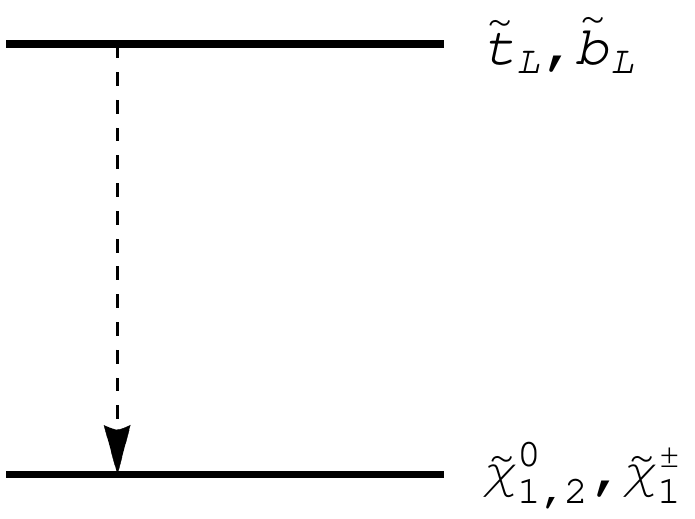}
\caption{ $\tilde t_L/\tilde b_L$-Higgsino}  \label{THtopology}
\end{subfigure}\hspace{.5cm}
\begin{subfigure}[b]{0.3\textwidth}\centering
\includegraphics[height=3.cm]{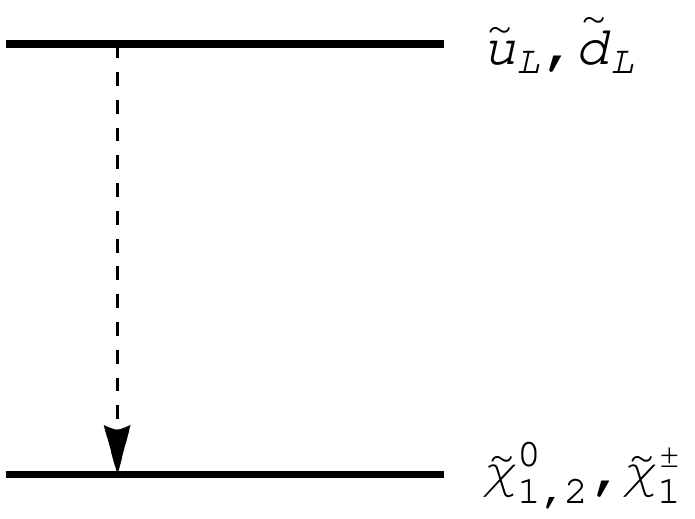}
\caption{$\tilde q_L$-Higgsino}  \label{qLHtopology}
\end{subfigure}\\
\vspace{0.3cm}
\begin{subfigure}[b]{0.3\textwidth}\centering
\includegraphics[height=3.cm]{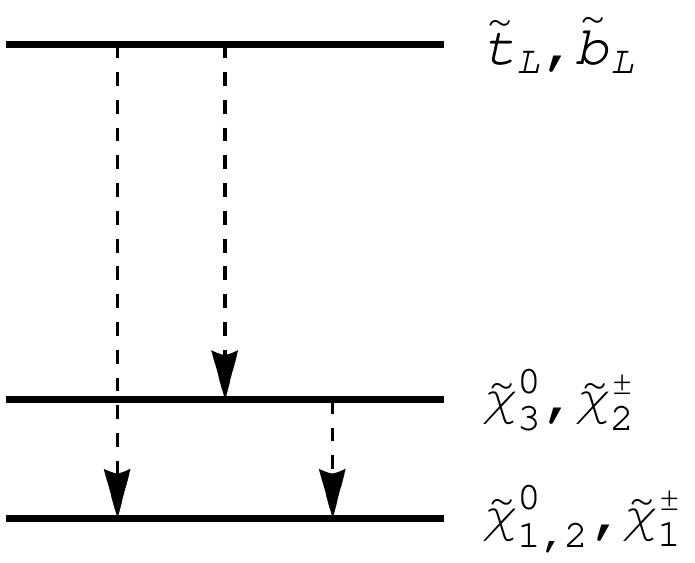}
\caption{$\tilde t_L/\tilde b_L$-wino-Higgsino }  \label{TWHtopology}
\end{subfigure}\hspace{3cm}
\begin{subfigure}[b]{0.3\textwidth}\centering
\includegraphics[height=3.cm]{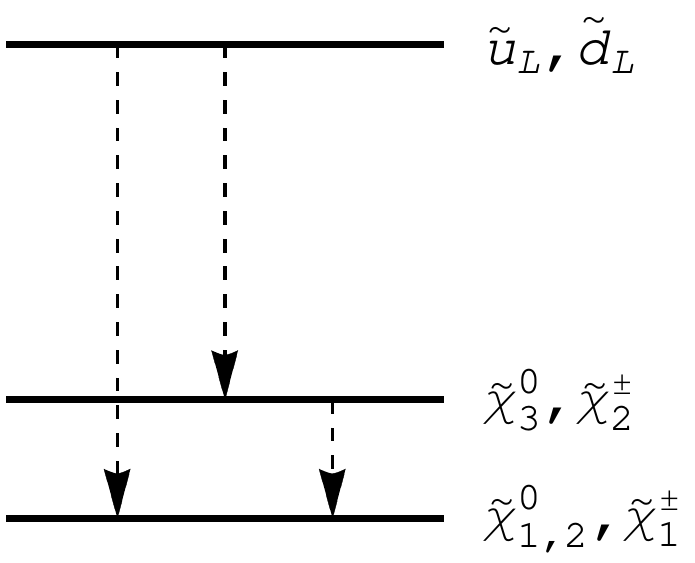}
\caption{$\tilde q_L$-wino-Higgsino }  \label{qLWHtopology}
\end{subfigure}
\caption{Simplified topologies for a Higgsino NLSP.}\label{fig:Higgsinotopo}
\end{figure}

As for the wino NLSP the bounds from monojet+MET searches are very weak, and are expected to probe Higgsino only up to $\sim 200$ GeV at HL-LHC \cite{Low:2014cba,Han:2013usa}. The remaining simplified topologies for the wino NLSP also apply for the Higgsino NLSP, with the roles of the wino and Higgsino interchanged. (See figs.~\ref{WHtopology}, \ref{THtopology}, \ref{qLHtopology} and \ref{TWHtopology}.) For $\tilde q_L$-Higgsino topology the analysis is identical as for the corresponding case with a wino NLSP and we assume the same limits. The decay in  the $\tilde t_L/\tilde b_L$-Higgsino topology is dominated by the top yukawa and the final states are therefore predominantly top-rich. In particular, for $\tan\beta=20$ the branching ratio to $t\bar t$+MET is roughly 90\% for both $\tilde t_L$ and $\tilde b_L$. Rescaling the (projected) limits in \cite{Khachatryan:2016xvy} and \cite{ATL-PHYS-PUB-2013-011} we find $m_{\tilde t_L,\tilde b_L}\gtrsim 850$ GeV and $m_{\tilde t_L,\tilde b_L}\gtrsim 1450$ GeV for the current and HL-LHC limits respectively.

\begin{figure}[h]
\includegraphics{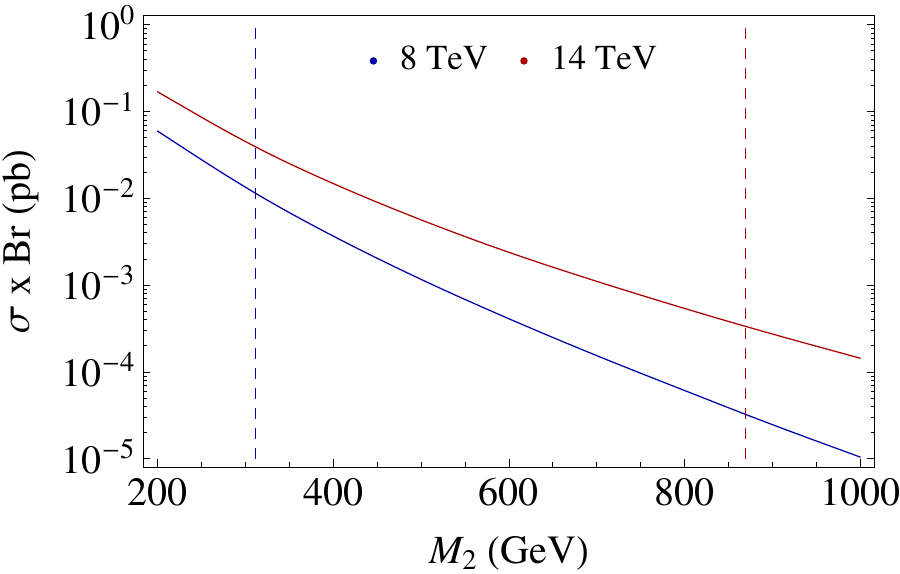}
\caption{Effective rate for the $WZ$+MET final state from wino production with a Higgsino NLSP. The dashed blue (red) lines indicate the current (projected) limits.\label{fig:HiggsinoWZplot}}
\end{figure}

\begin{figure}[t]
\begin{floatrow}
\ffigbox{%
\includegraphics[width=0.45\textwidth]{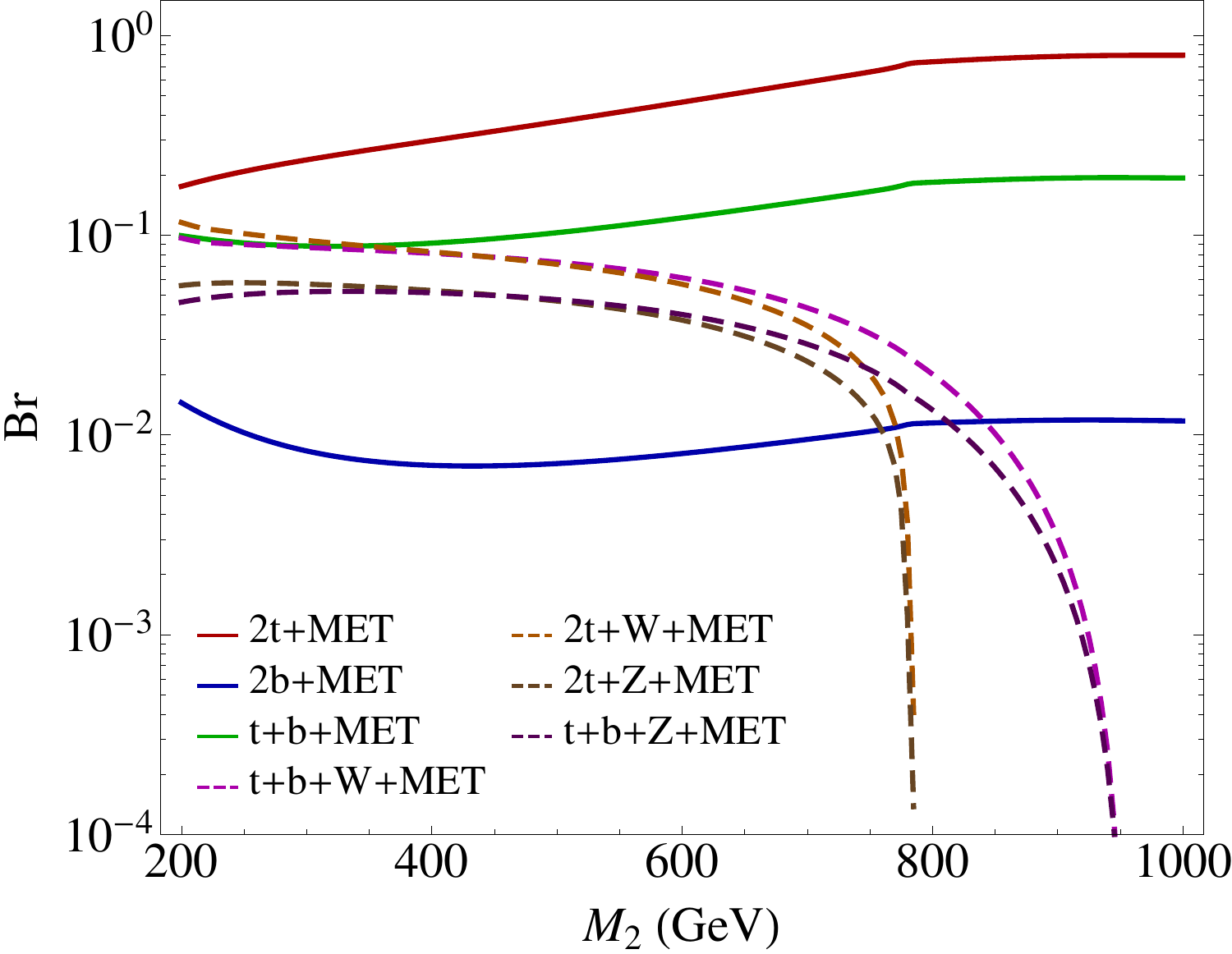}%
}{%
 \caption{Branching ratios for some of the channels in the $\tilde t_L/\tilde b_L$-wino-Higgsino  topology, with $m_{\tilde t_L}=m_{\tilde b_L}=1$ TeV and $\mu=100$ GeV.\label{fig:TWHcascade}}%
}\hfill
\ffigbox{%
\includegraphics[width=0.45\textwidth]{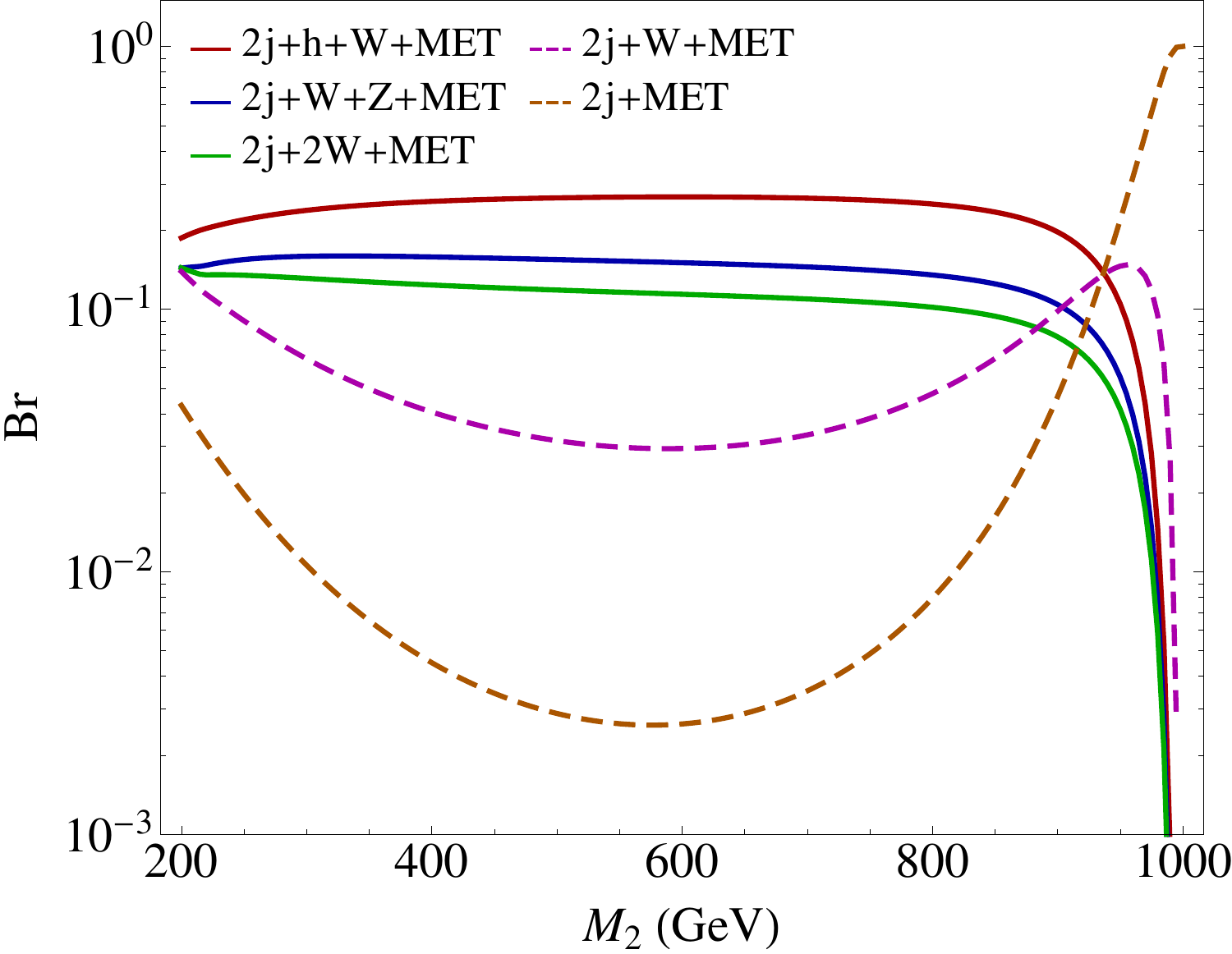}%
}{%
  \caption{Branching ratios for some of the channels in the $\tilde q_L$-wino-Higgsino  topology, with $m_{\tilde q_L}=1$ TeV and $\mu=100$ GeV.\label{fig:qLWHcascade}}%
}
\end{floatrow}
\end{figure}

 For the wino-Higgsino topology, which is relevant only for $\mu>0$, the branching ratios are modified with respect to the Higgsino-wino topology in fig.~\ref{HWtopology}. This leads to a different effective rate for $WZ+$MET final state\footnote{It is possible that the sleptons sit below the wino, which enhances the sensitivity, since a cascade decay with additional leptons in the final state becomes possible. This however happens only in a relatively small part of the parameter space, and we do not consider this possibility separately. }, as shown in fig.~\ref{fig:HiggsinoWZplot}. Similarly, the branching ratios for the $\tilde t_L/\tilde b_L$-wino-Higgsino are modified with respect to its analogue for with a wino NLSP (see fig.~\ref{fig:TWHcascade}). Finally, for the Higgsino NLSP there is a $\tilde q_L$-wino-Higgsino topology (fig.~\ref{qLWHtopology}), whose analogue was not needed for the wino NLSP. In this topology, the cascade decay via the wino always dominates, unless the wino is very close in mass to the squarks (see fig.~\ref{fig:qLWHcascade}). Also here, since the wino is always heavy for $\mu<0$, topologies \ref{WHtopology}, \ref{TWHtopology} and \ref{qLWHtopology} only apply if $\mu>0$. As for the wino NLSP, we do not attempt a proper recasting of the existing limits in terms of the cascade topologies of the colored sparticles, and instead we take the same squarks limits as for the wino NLSP. Especially for the squarks, the limits are expected to be very sensitive to the wino mass, as was shown in \cite{Aad:2015iea} for a bino NLSP. However when we marginalize over the full parameter space, as in fig.~\ref{villadoroH}, our crude approximation should give nevertheless a reasonable idea of the overall strength and relative importance of the various limits.

\subsection{Bino NLSP} 
\begin{figure}[t]\centering
\includegraphics[width=\textwidth]{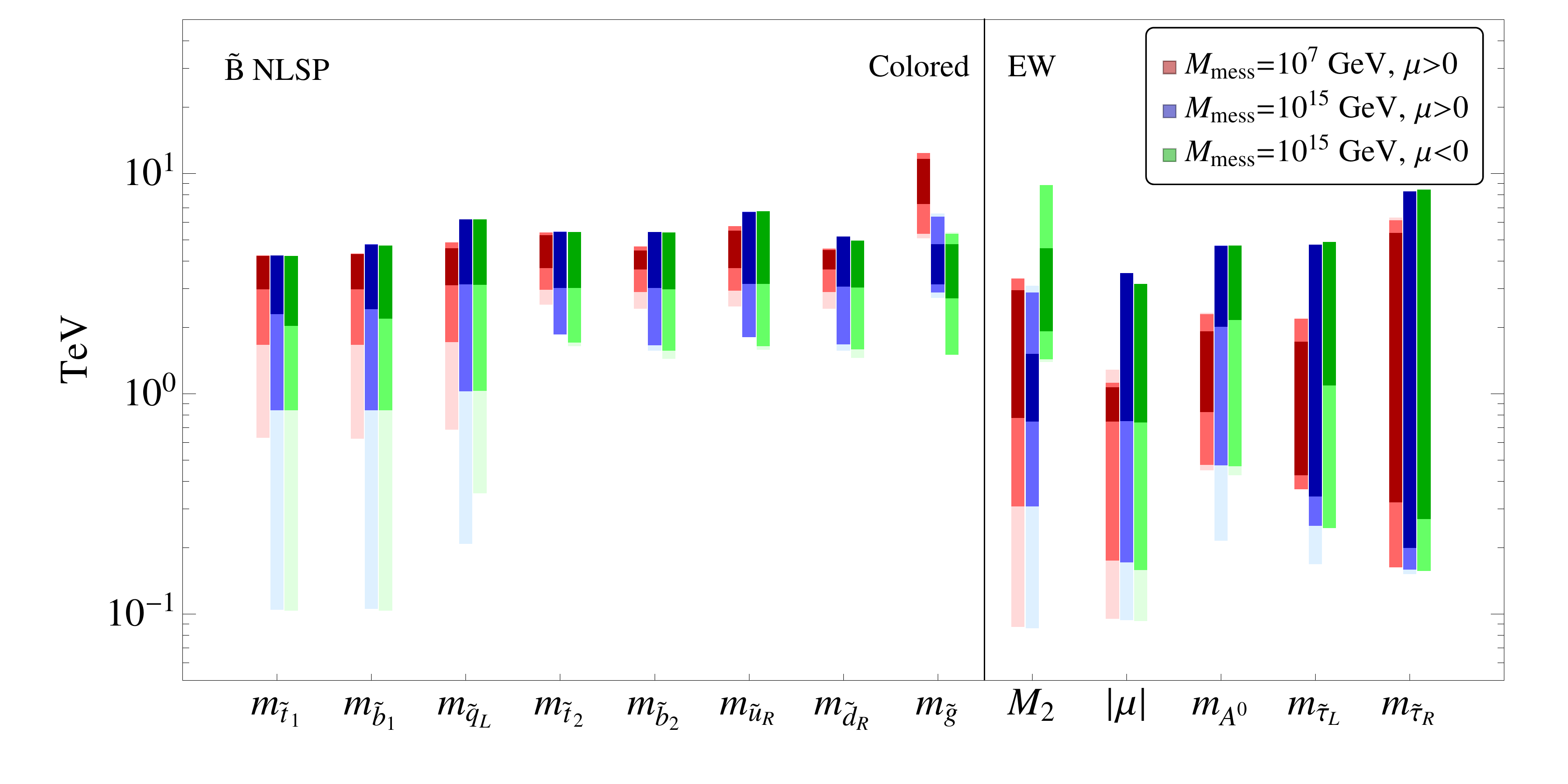}
\caption{Overview of the GGM parameter space for a bino NLSP. Regions with lightest shading are excluded or disfavored by existing data, regions with darker shading are accessible at the HL-LHC, while darkest regions are likely to be unaccessible at the LHC.}  \label{villadoroB}
\end{figure}

The bino NLSP scenario, summarized in fig.~\ref{villadoroB}, is considerably simpler than the wino and Higgsino NSLP: While the wino and Higgsino masses strongly correlate with the rest of the spectrum, the bino mass has essentially no impact. This also implies that the parameter space for the bino NLSP is substantially larger than for the Higgsino and the wino NLSP, since it is always possible to put the bino at the bottom of the spectrum. More importantly, the bottom of the spectrum then consists out of single neutralino, rather than a quasi degenerate multiplet of several electroweakinos. This greatly simplifies the analysis of the branching ratios in most of the simplified topologies. For this reason simplified topologies with bino (N)LSP have been a popular method to parametrize the experimental limits, and for most topologies considered here a direct limit is available.

\begin{figure}[h!]\centering
\includegraphics[width=0.5\textwidth]{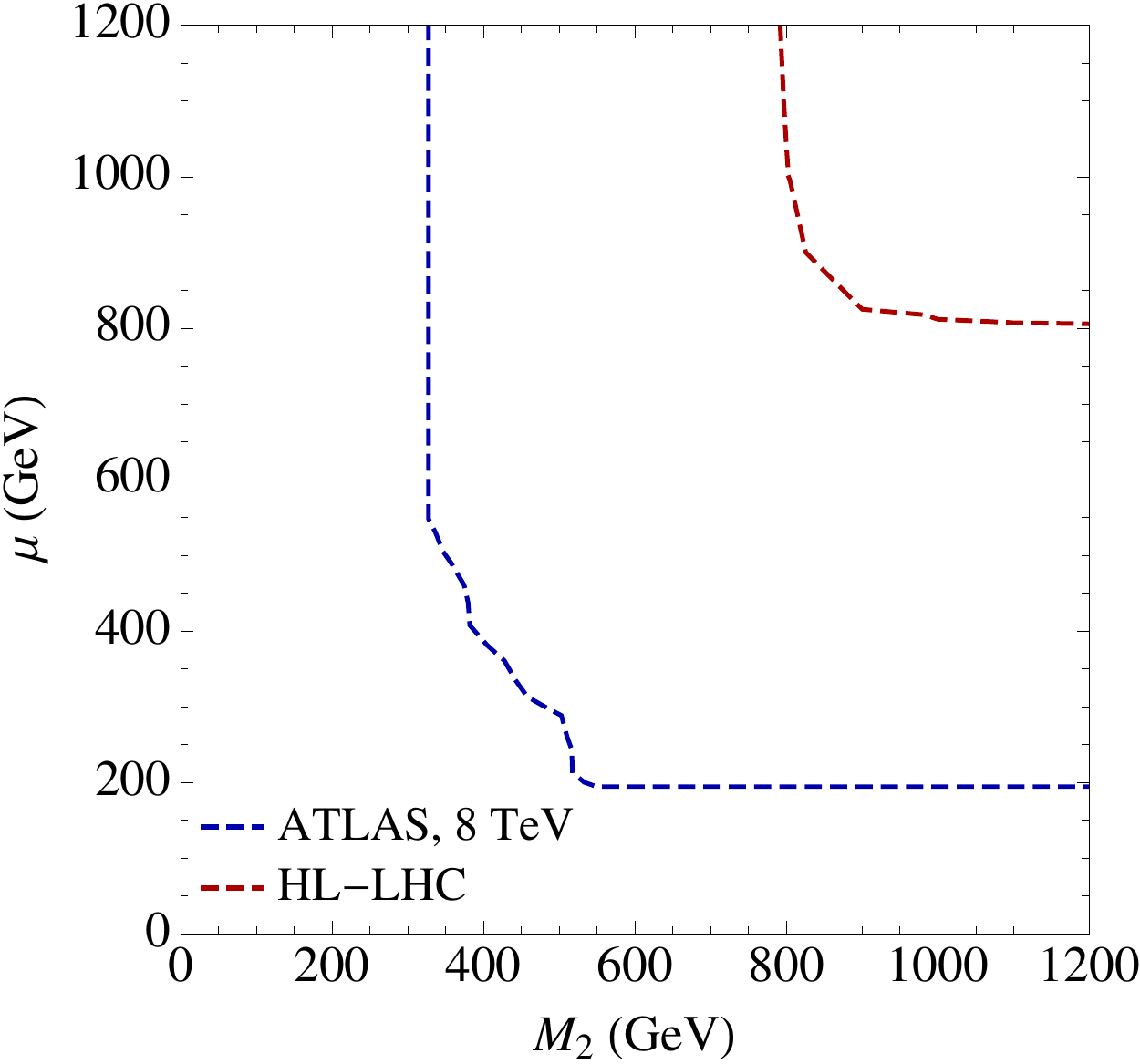}
\caption{ Current limit (blue) \cite{Aad:2014vma} and estimated reach (red) on the wino-Higgsino-bino simplified topology with $M_1=50$ GeV. The projected limit was rescaled from \cite{ATL-PHYS-PUB-2014-010}. }  \label{BinoNLSPlimits}
\end{figure}

Again neglecting the sleptons, the relevant simplified topology for electroweak production simply consists of the wino and/or Higgsino in addition to the bino NLSP. ATLAS has set a limit directly on this topology in the $M_2$-$\mu$ plane \cite{Aad:2014vma}, where the strongest limit again comes from the $WZ$+MET final state, as shown in fig.~\ref{BinoNLSPlimits}. In fig.~\ref{BinoNLSPlimits} we also estimate the reach for the HL-LHC, where we rescale the projection for the wino-bino simplified topology in \cite{ATL-PHYS-PUB-2014-010} to also allow for the possibility of an accessible Higgsino.

For colored production the topologies are the same as those for the Higgsino NLSP, upon replacing the Higgsino multiplet with the bino. In addition to the trivial $\tilde q_L$-bino topology, the $\tilde q_L$-wino-bino topology is relevant as well. To estimate the current and projected bounds, we follow the same procedure as for wino and Higgsino NLSP. For the $\tilde q_L$-wino-bino model the limit in principle varies significantly with the wino mass \cite{Aad:2015iea}, but when we marginalize over the parameter space in fig.~\ref{villadoroB} our simple minded procedure still provides a reasonable idea of the impact of the bound on the GGM parameter space.

 The limits on the $\tilde t_L/\tilde b_L$-bino topology can be read off directly from the stop and sbottom searches. The strongest (projected) limit is $m_{\tilde t_L,\tilde b_L}\gtrsim900$ GeV \cite{ATLAS-CONF-2015-066,CMS-PAS-SUS-16-001} ($m_{\tilde t_L,\tilde b_L}\gtrsim$ 1500 GeV \cite{ATL-PHYS-PUB-2014-010}), where we again took the left-handed stop and sbottom to be quasi degenerate. In a significant part of the parameter space it is moreover possible that both the wino and the Higgsino are in between the stop/sbottom and the bino NLSP. In this case a very complicated, multi-step cascade is possible. Finally, for the HL-LHC reach of the gluino, we employ the same estimate as for the Higgsino NLSP.

\subsection{Sneutrino NLSP} 
The sneutrino NLSP behaves qualitatively similar to the neutralino NLSPs, except that the decay chains tend to be lepton and/or tau rich, which can result in a very rich and interesting phenomenology \cite{Figy:2010hu,Katz:2009qx,Katz:2010xg,Arina:2013zca,Chatterjee:2016rjo}.
Although the snutau is typically slightly lighter than the remaining sneutrino's, we can take all three flavors as degenerate for the purpose of this discussion. The mass splitting between $\tilde \tau_L$ and $\tilde \nu_\tau$ is typically 5 GeV or less, which means that any leptons produced in the $\tilde \tau_L\rightarrow W^\ast \tilde \nu_\tau$ decay tend to be fairly soft. This is especially so for direct $\tilde \tau_L^+\tilde \tau_L^-$ production, which is usually produced on or near threshold. Direct production of the sneutrino NLSP is therefore notoriously difficult to constrain at the LHC, and the presence of heavier states with larger cross sections is therefore essential to probe this scenario efficiently. 

\begin{figure}[t]\centering
\includegraphics[width=1\textwidth]{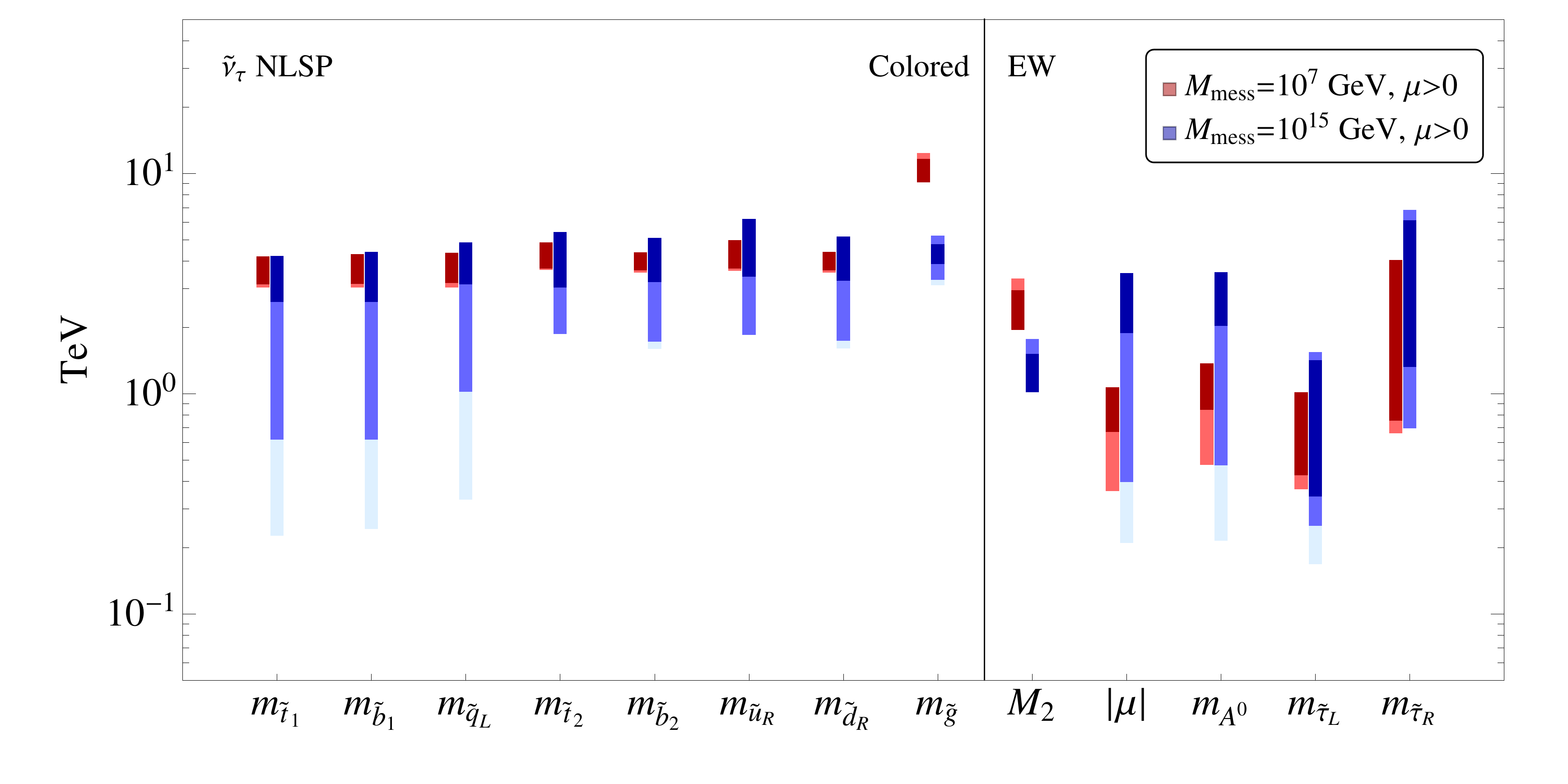}
\caption{Overview of the GGM parameter space for a snutau NLSP. Regions with lightest shading are excluded or disfavored by existing data, regions with darker shading are accessible at the HL-LHC, while darkest regions are likely to be unaccessible at the LHC.  $M_{mess}=10^{15}$ GeV with $\mu<0$ is not included, see text for details.  The (projected) bounds on the colored sfermions are to be understood as very rough estimates only, see text for details.\label{villadoronuTau}}
\end{figure}

\begin{figure}[t]\centering
\begin{subfigure}[b]{0.3\textwidth}\centering
\includegraphics[height=3.cm]{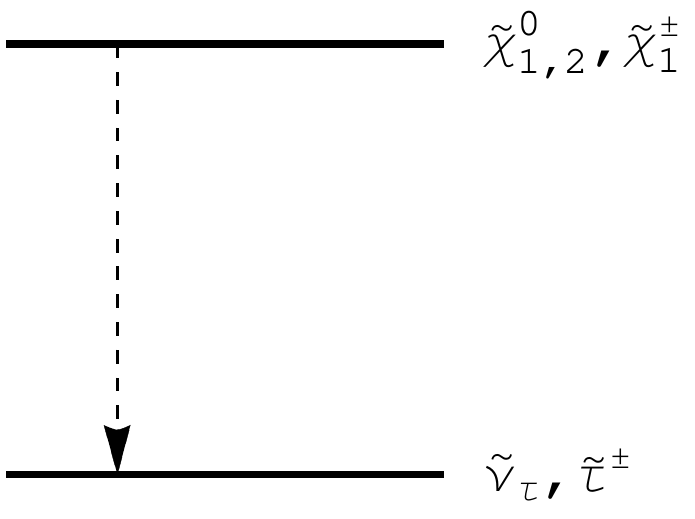}
\caption{ Higgsino-sneutrino}  \label{Hnutopology}
\end{subfigure}\hspace{.5cm}
\begin{subfigure}[b]{0.3\textwidth}\centering
\includegraphics[height=3.cm]{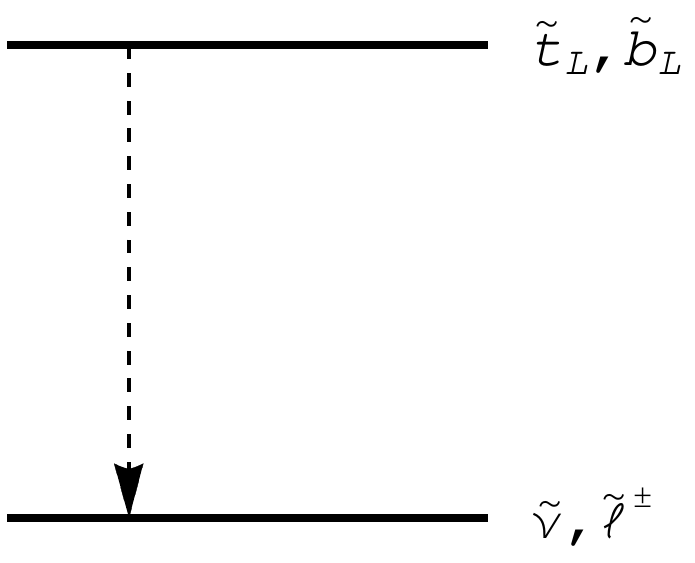}
\caption{ $\tilde t_L/\tilde b_L$-sneutrino}  \label{Tnutopology}
\end{subfigure}\hspace{.5cm}
\begin{subfigure}[b]{0.3\textwidth}\centering
\includegraphics[height=3.cm]{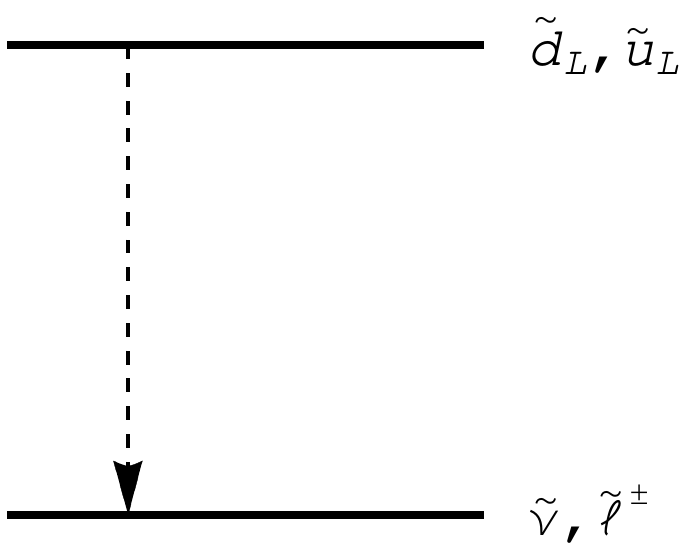}
\caption{$\tilde q_L$-sneutrino}  \label{qLnutopology}
\end{subfigure}\\
\vspace{0.3cm}
\begin{subfigure}[b]{0.3\textwidth}\centering
\includegraphics[height=3.cm]{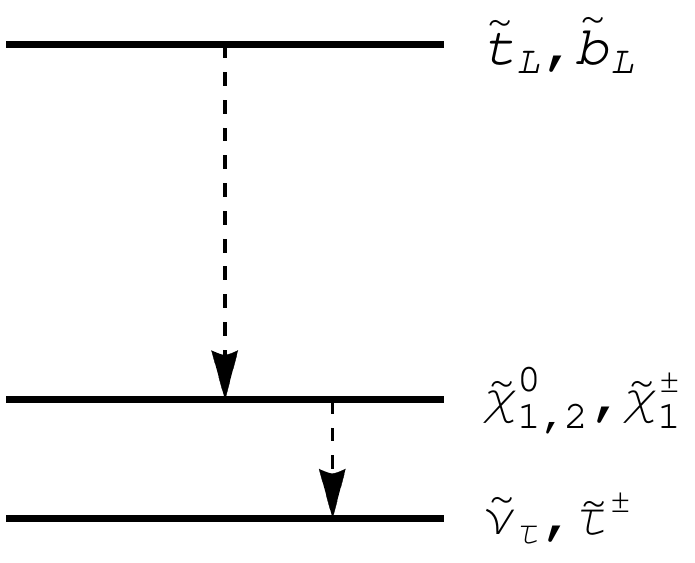}
\caption{$\tilde t_L/\tilde b_L$-Higgsino-sneutrino}  \label{THnutopology}
\end{subfigure}\hspace{3cm}
\begin{subfigure}[b]{0.3\textwidth}\centering
\includegraphics[height=3.cm]{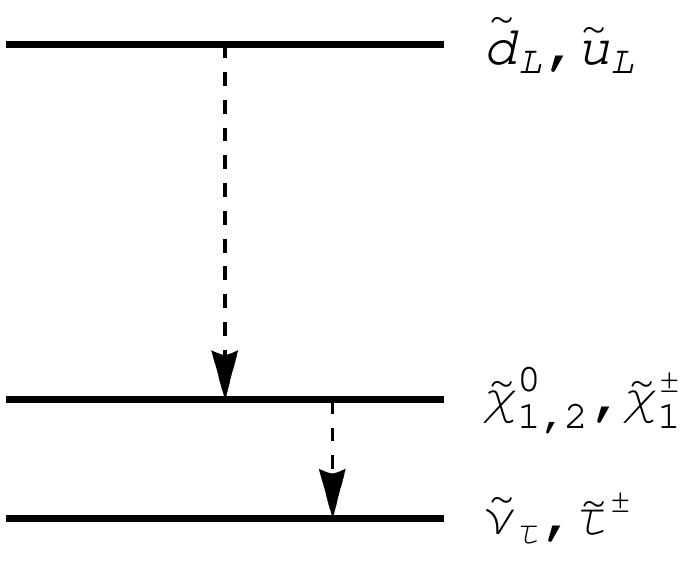}
\caption{$\tilde q_L$-Higgsino-sneutrino}  \label{qLHnutopology}
\end{subfigure}
   \caption{Simplified topologies for a sneutrino NLSP. $\tilde \nu$ and $\tilde \ell$ stand for all three generation sleptons. For the topologies indicated with $\tilde \nu_\tau, \tilde \tau^\pm$, all generation sleptons are in principle present, but only the third generation contributes significantly to phenomenology.}\label{fig:sneutrinotopo}
\end{figure}

The sneutrino NLSP can only occur when $m_{Q_3}<m_{U_3}$, as shown in fig.~\ref{NLSPplots}. Moreover for $M_{mess}=10^{15}$ GeV with $\mu<0$ a large amount of tuning is needed to achieve a sneutrino NLSP and EWSB simultaneously, and a specialized study is therefore required to make quantitatively accurate statements about this scenario. Since this concerns a very small part of the GGM parameter space, we do not attempt this here. For $M_{mess}=10^{15}$ GeV with $\mu>0$, a sneutrino NLSP is more likely, but even in this case the remainder of the spectrum is very constrained by the EWSB conditions and the GGM sum rules: Aside from the $m_{Q_3}<m_{U_3}$ requirement, we find that the wino mass is always in the (multi-)TeV range, as shown in fig.~\ref{villadoronuTau}. This means that it is irrelevant for direct production and for on-shell cascade decays, such that the relevant simplified topologies are just those shown in fig.~\ref{fig:sneutrinotopo}. 

For $M_{mess}=10^{7}$ GeV the spectrum is even more restricted, and the only relevant topology is the Higgsino-sneutrino topology in fig.~\ref{Hnutopology}. For this topology, the most promising channel is $2\tau+$MET, which has branching fraction 1/4. However even in this channel the SM background from $WW$ production is very large, and there is currently no limit from the LHC once the LEP bound on $\tilde \nu$ is accounted for \cite{Aad:2014yka,CMS-PAS-SUS-14-022}. (It is however possible that mild bounds could be obtained by accounting for possible soft leptons in the $\tilde \tau_L$ decay.) In particular for low scale gauge mediation we thus expect that it will be very difficult for the LHC to significantly constrain the sneutrino NLSP scenario in this way. On the other hand, the CP-odd Higgs tends to be light if the Higgsino is light (see eqn.~\eqref{IRexact}). As for the Higgsino NLSP, this effect is more pronounced for low messenger scales and therefore provides an interesting, complementary constraint of $m_{A^0}\gtrsim 500$ GeV from current data \cite{CMS:2015mca,Aad:2014vgg}. For the HL-LHC we obtain $m_{A^0}\gtrsim 900$ GeV by rescaling the projection in \cite{Djouadi:2015jea}.

For $M_{mess}=10^{15}$ GeV with $\mu>0$, the left-handed squarks, stop and sbottom can be accessible with current data, which could provide a much needed boost to the signal cross section. An important difference with the neutralino NSLP is that the squarks, stop and sbottom cannot directly decay to the NLSP, but must go through an intermediate electroweakino. This intermediate electroweakino can be either an on-shell Higgsino (figs.~\ref{THnutopology} and \ref{qLHnutopology}) or an off-shell wino and/or Higgsino (figs.~\ref{Tnutopology} and \ref{qLnutopology}). The on-shell cascade takes priority whenever it is available.

Since the wino is heavy, the on-shell decays must go through the Higgsino and these decay chains are therefore characterized by (nu)tau-rich final states with missing energy in addition to tops, results or jets. A particularly interesting mode is $\tilde t_L\tilde t_L$ to a pair of on-shell Higgsinos, which predominantly decays in $t\bar t \tau^\pm\tau^\mp$+MET and $t\bar t \tau^\pm\tau^\pm$+MET with roughly 40\% branching ratio each. This decay topology is currently not explicitly covered, although it should be possible to regain some sensitivity by recasting the multi-lepton searches or the $2b+\tau^+\tau^-$ + MET search \cite{Aad:2015tin}. Since the latter search is inclusive as far as the number of jets is concerned, we can obtain $m_{\tilde t_L}\gtrsim 500$ GeV as a naive estimate of the bound by rescaling the observed bound to account for the branching ratio $2t+\tau^+\tau^-$, with fully hadronic top decays. The off-shell, three-body squark decay  in fig.~\ref{qLnutopology} prefers an off-shell wino and therefore tends to be lepton flavor democratic. For the stop/sbottom initiated  three-body decay in fig.~\ref{Tnutopology}, the branching ratios are sensitive to the relative masses of the wino and the Higgsino. In this paper we do not attempt to extract an approximate bound for this case.

In summary, the phenomenology of $\tilde t_L/\tilde b_L/\tilde q_L$ is very rich and complicated: Some channels are currently not or poorly covered, while others require detailed recasting \cite{Guo:2013asa}. To get very rough, qualitative sense of what impact of such a program would be on the GGM parameter space, we imposed the same (projected) squark bounds as for the Higgsino NLSP in fig.~\ref{villadoronuTau}, in addition our naive rescaling of the $2b+\tau^+\tau^-$ + MET search. The true bounds are likely to be stronger due to the additional leptons and/or tau's in most of the channels.

\section{Charged/colored NLSP}\label{sec:ch/colNLSP}

We again assume the NLSP to be long-lived, something which is preferred for $M_{mess}\gtrsim 10^7$ GeV. For $M_{mess}\lesssim 10^{7}$ GeV prompt or displaced decays are possible, however as we will see, this case does not allow for a stop/sbottom NLSP. A promptly decaying stau NLSP is possible, and in this case the limits are very weak since colored production is not accessible for low $M_{mess}$. (See for \cite{Ruderman:2010kj,Kats:2011qh} for a discussion of electroweak production with a promptly decaying stau NLSP.) For a discussion of the displaced decay of a stau NLSP we refer to \cite{Asai:2011wy,Evans:2016zau}.

For a long-lived stau, stop and sbottom NLSP's the situation is considerably simpler than for a neutral NLSP. While there can be some efficiency loss for cascades with particularly large hierarchies or with heavy NLSPs \cite{Heisig:2015yla}, generally the details of the spectrum are much less important than for searches which rely on MET. The searches for heavy stable charged particles (HSCP's) can therefore be interpreted as a rough bound on total inclusive sparticle cross section. The current bounds and projected limits are summarized in tab.~\ref{chargedNLSP}. To estimate the current and future bounds on the parameter space we compute the total SUSY cross section by adding up the individual pair production cross sections of each sparticle and compare this with the corresponding limit in tab.~\ref{chargedNLSP}.  

\begin{table}[h]
\begin{tabular}{|c|cc|cc|cc|}\hline
&\multicolumn{2}{c|}{8 TeV}  &\multicolumn{2}{c|}{13 TeV}   & \multicolumn{2}{c|}{HL-LHC}\\\hline
$\tilde t$ NLSP&1 fb&\cite{Chatrchyan:2013oca,ATLAS:2014fka}&10 fb&\cite{CMS-PAS-EXO-15-010}&$1\times 10^{-2}$ fb&\cite{Gershtein:2013iqa}\\
$\tilde \tau$ NLSP&0.3 fb&\cite{Chatrchyan:2013oca,ATLAS:2014fka}&2 fb &\cite{CMS-PAS-EXO-15-010}&$2\times 10^{-3}$ fb&\cite{Gershtein:2013iqa}\\\hline
\end{tabular}
\caption{Existing limits and projected exclusion power on the inclusive cross section for HSCP's. \label{chargedNLSP}}
\end{table}

\subsection{Slepton NLSP}

The stau NLSP scenario can be realized only if $m_{Q_3}>m_{U_3}$, which implies that colored production is very suppressed due to the strong lower bound on the right-handed squarks (see fig.~\ref{hwplots}). The cross section is therefore dominated by electroweak production of electroweakinos (Higgsinos in all cases and winos if $\mu>0$), while slepton pair production is subdominant.

\begin{figure}[p]\centering
\includegraphics[width=1\textwidth]{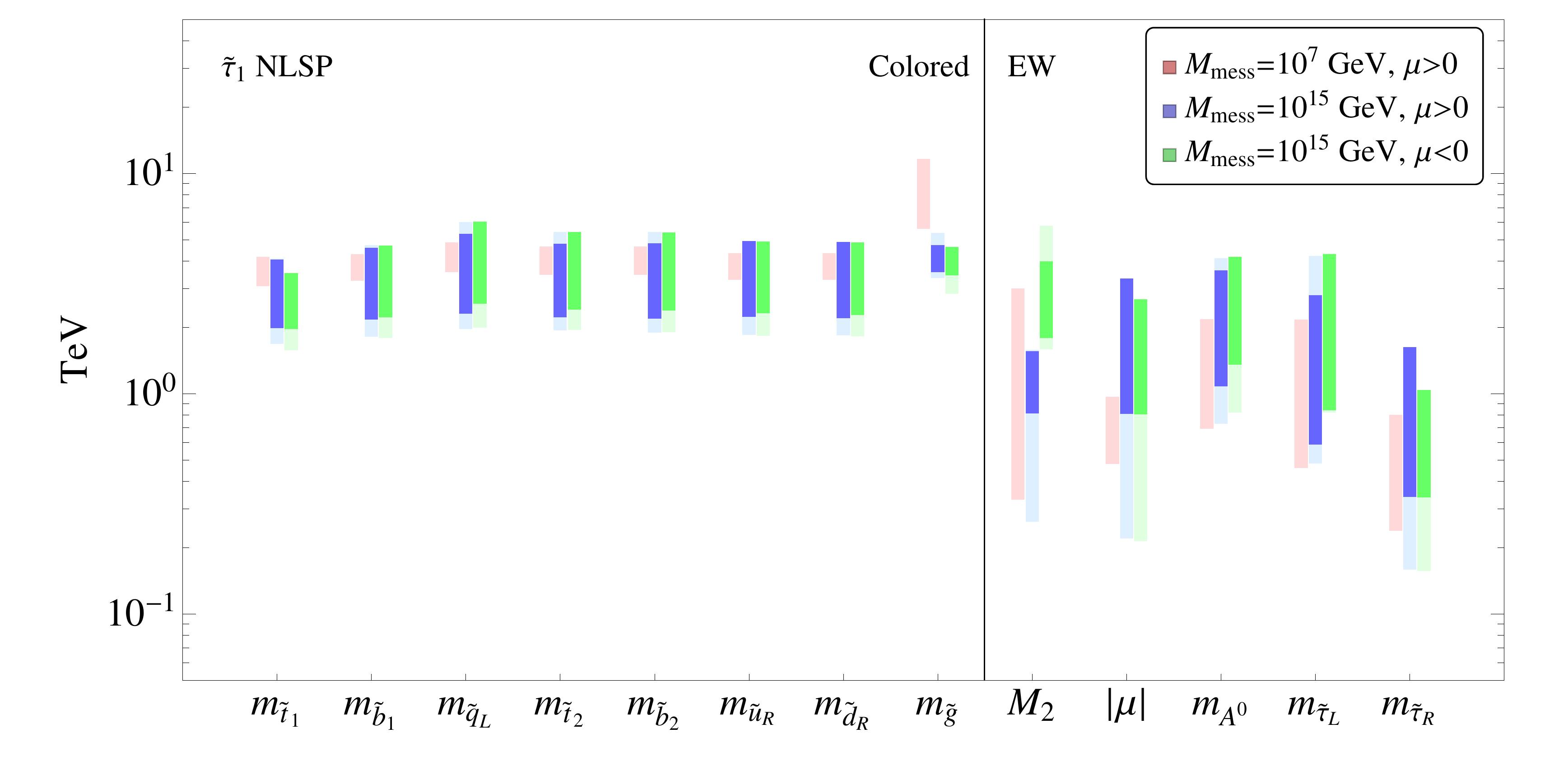}
\caption{Overview of the GGM parameter space for a stau NLSP. Regions with lightest shading are excluded or disfavored by existing data, regions with darker shading are accessible at the HL-LHC. }  \label{villadoroTau}
\end{figure}

The present bounds from the LHC are already severely constraining the parameter space with a stau NLSP, pushing the mass of the lightest chargino $\chi_1^{\pm}$ to be heavier than $\sim900\text{ GeV}$ for $M_{mess}=10^{15}$ GeV. For $M_{mess}=10^{7}$ GeV both $M_2$ and $\mu$ are both forced to be light (see fig~\ref{hwplots}), which increases the electroweak cross section relative to $M_{mess}=10^{15}$ GeV. As a result, the stau NLSP for $M_{mess}=10^{7}$ GeV is already excluded by the present LHC data. The remainder of the stau NLSP parameter space for $M_{mess}=10^{15}$ GeV can be probed at the HL-LHC.

\subsection{Stop/sbottom NLSP}

\begin{figure}[p]\centering
\includegraphics[width=1\textwidth]{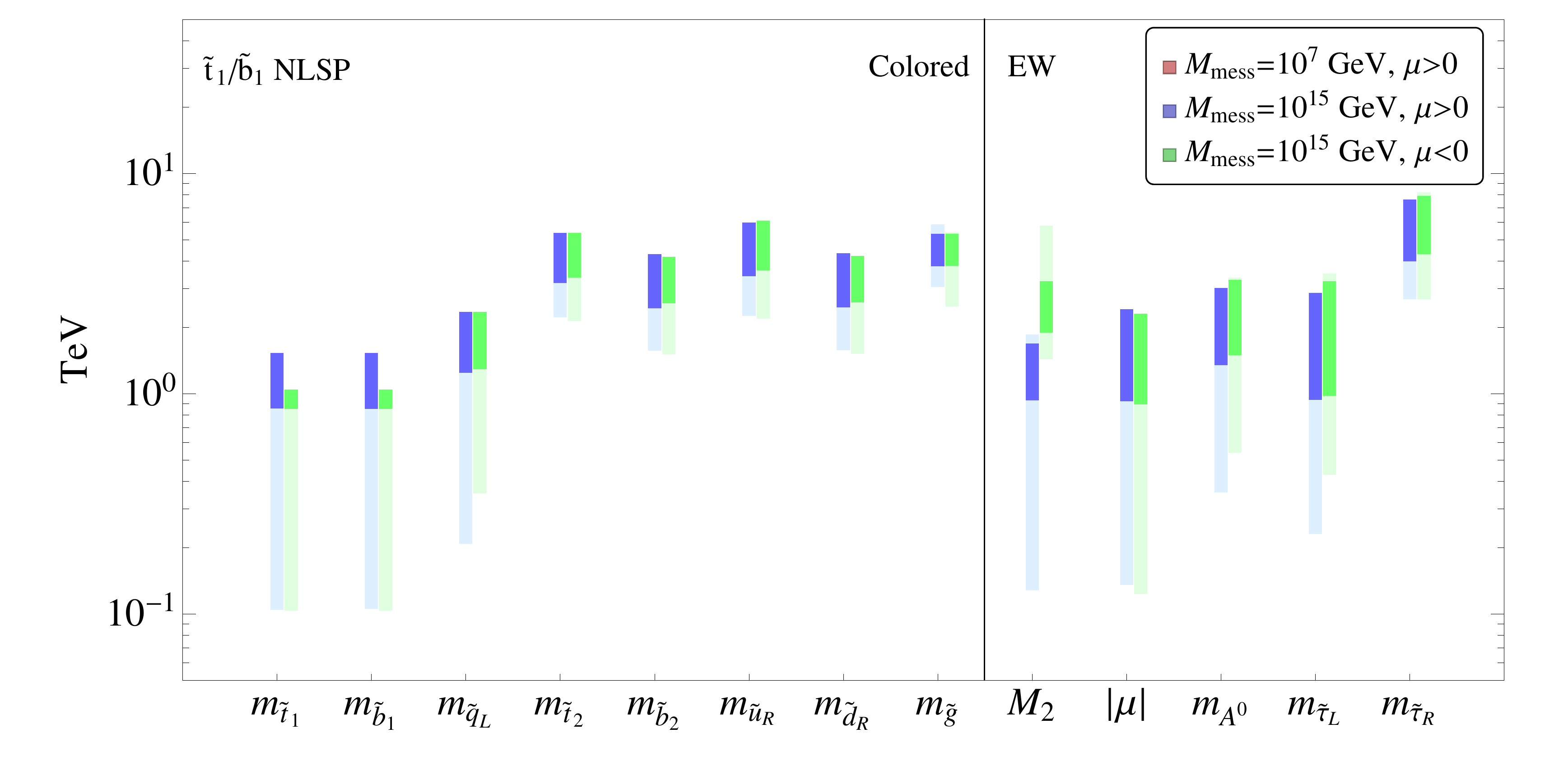}
\caption{Overview of the GGM parameter space for a stop/sbottom NLSP. Regions with lightest shading are excluded or disfavored by existing data, regions with darker shading are accessible at the HL-LHC. There is no allowed parameter space for $M_{mess}=10^7$ GeV.}  \label{villadoroT}
\end{figure}

Stop or sbottom can only be the NLSP if $M_{mess}=10^{15}$ GeV, in which case one expects the NLSP to be long-lived. Conversely to the stau NLSP case, direct production of NLSP pairs dominates the SUSY cross section almost everywhere in the parameter space. The present and projected constraints are shown in fig.~\ref{villadoroT}: the stop NLSP scenario is already very tightly constrained with the existing data, and will be fully probed at the HL-LHC.

\section{Summary and conclusions\label{sec:conclusions}}
Based on the solution obtained in \cite{Knapen:2015qba}, we presented a full characterization of the phenomenology of general gauge mediation with stops below 4 TeV. The Higgs mass constraint is hereby crucial, and dramatically reduces the freedom inherent in the GGM parameter space. In particular:
\begin{itemize}
\item The gluino is almost always decoupled at LHC, except for a small region of parameter space with $\mu<0$. The right-handed squarks are currently not accessible, but for high messenger scales they could be within the reach of the high luminosity LHC.
\item The Higgsino mass is bounded from above. For $\mu>0$ the wino mass is also bounded from above while for $\mu<0$, the wino mass is bounded from below.
\item The nature of the slepton NLSP is very tightly correlated with the nature of the lightest squarks, which determine the maximal colored production. While for a sneutrino NLSP the lightest squarks are left-handed and the colored production can be sizable, the lightest squarks are right-handed and strongly bounded from below if the NLSP is a stau. In the latter case only electroweak states dominate the SUSY cross section.
\item Higher messenger scales enlarge the allowed parameter space. As a result, long NLSP life-times are favored in most models. 
\end{itemize}

These features have a strong impact on the collider phenomenology: We find that both the stop/sbottom and stau NLSP scenarios are already very strongly constrained by the existing data. For low messenger scale these NLSP types are already excluded, while for higher messenger scale both scenarios will be probed completely at HL-LHC. For the neutral NLSP's (wino, Higgsino, bino and sneutrino) there is a relatively small number of simplified topologies (figs.~\ref{fig:Winotopo}, \ref{fig:Higgsinotopo} and \ref{fig:sneutrinotopo}) which span nearly all of the accessible parameter space. We employed those handful of topologies to estimate the limits and asymptotic reach for each NLSP type in figs.~\ref{villadoroW}, \ref{villadoroH}, \ref{villadoroB} and \ref{villadoronuTau}. 

\begin{figure}[t]\centering
\includegraphics[width=0.3\textwidth]{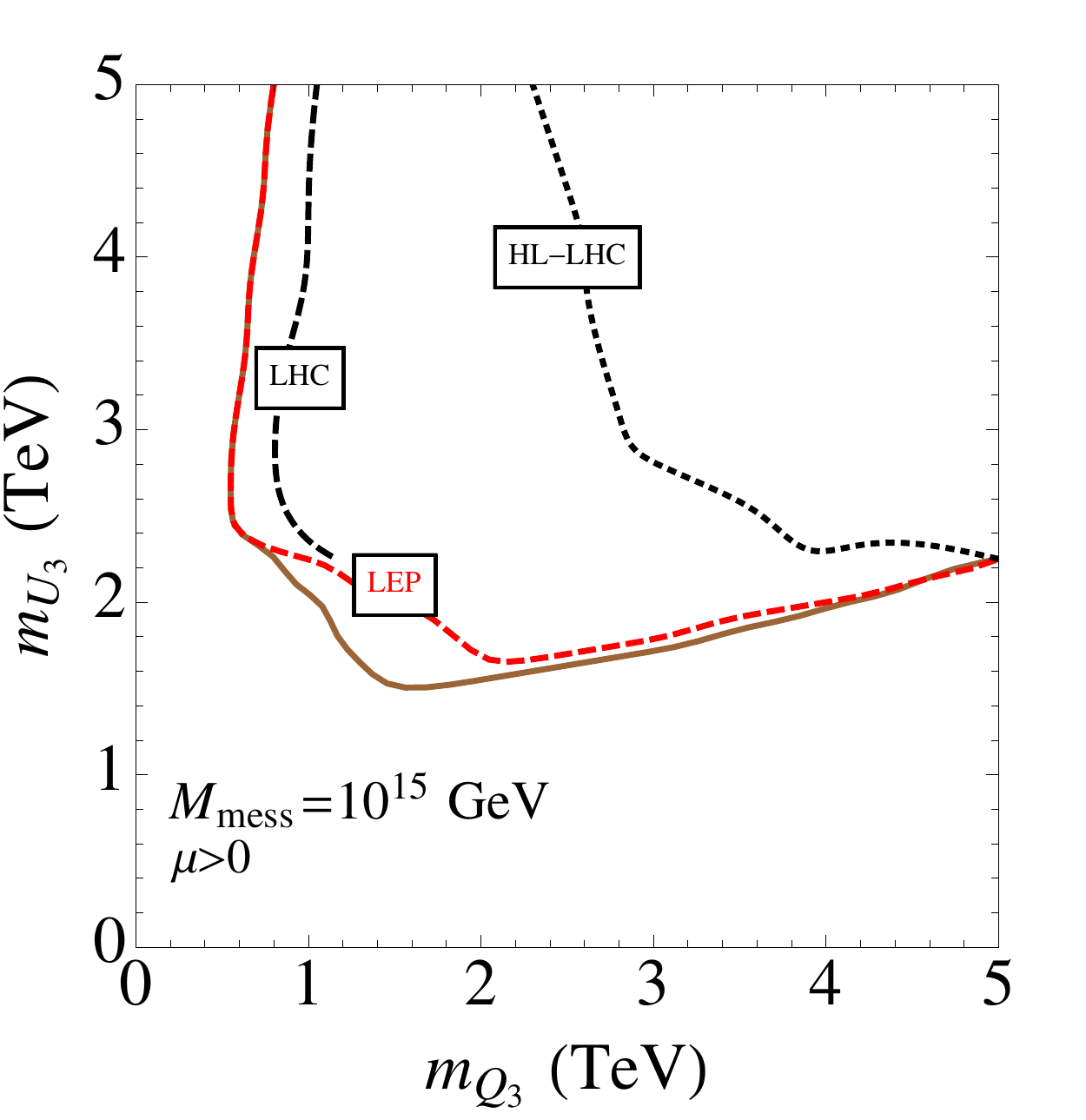}\hfill
\includegraphics[width=0.3\textwidth]{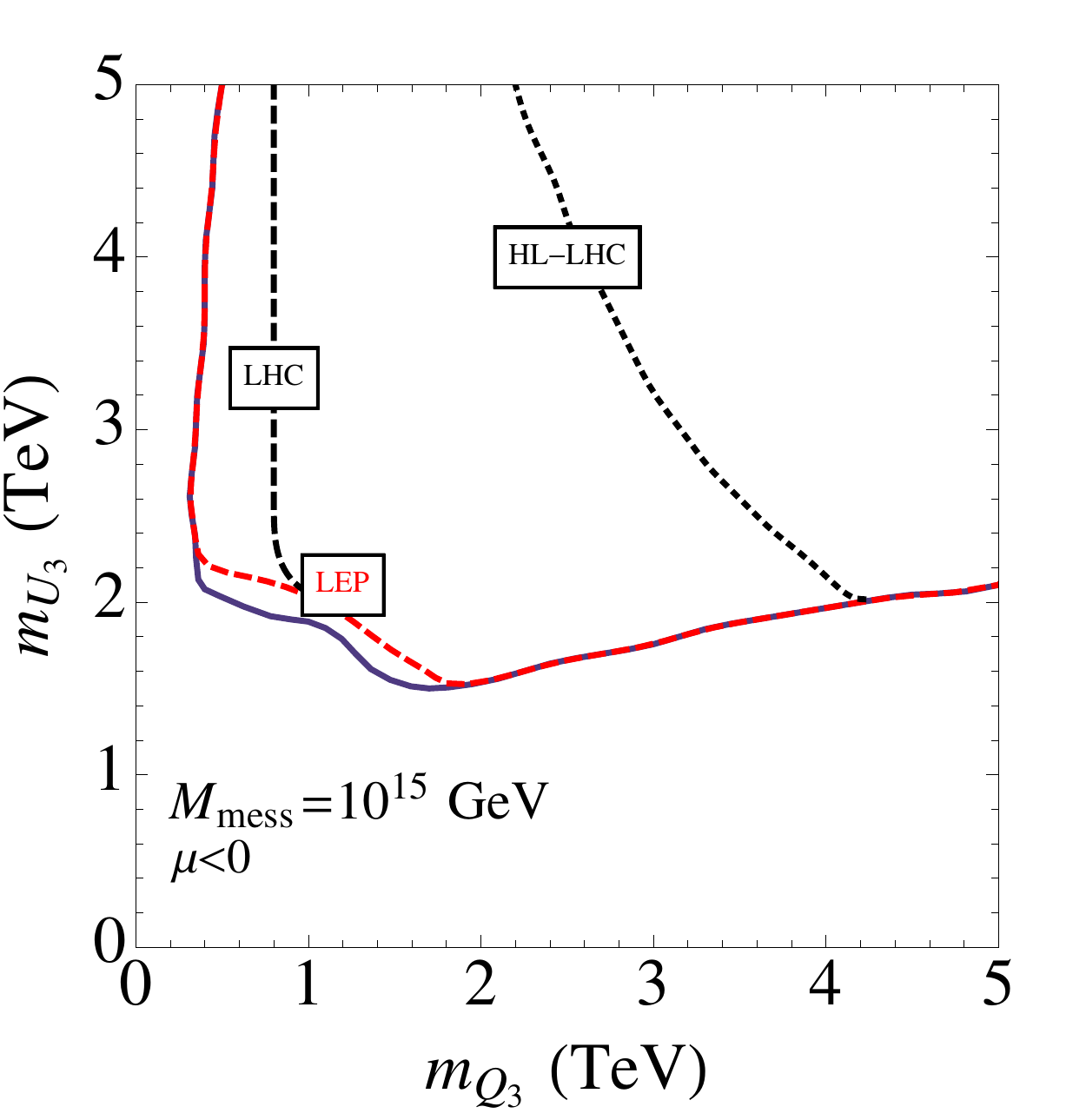}\hfill
\includegraphics[width=0.3\textwidth]{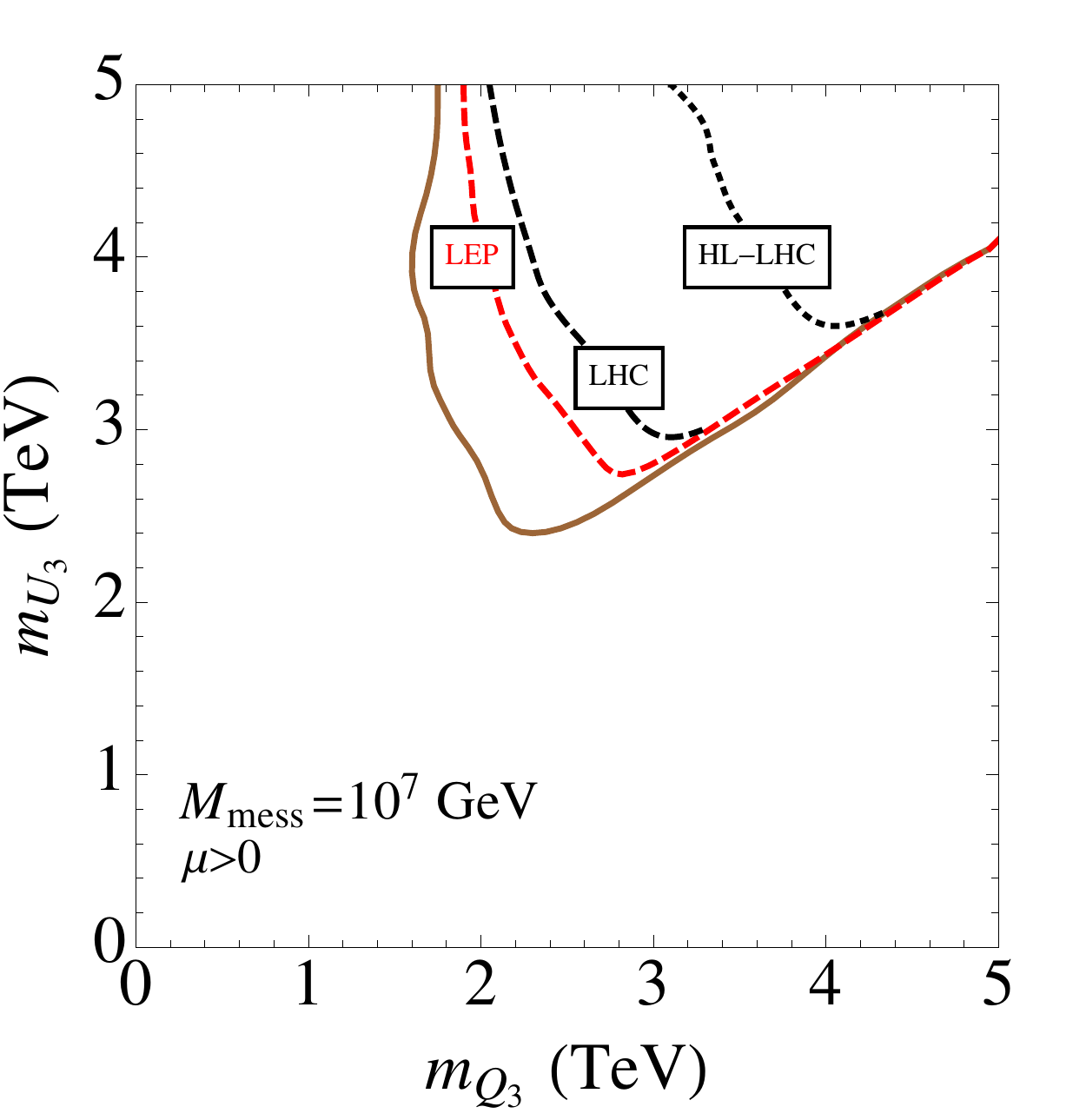}
\caption{Exclusion reach in the stop mass plane of LEP (red dashed) , current LHC data (black dashed) and HL-LHC with 3000 $\text{fb}^{-1}$  (black dotted). \label{conclusion_plot}}
\end{figure}

Fig.~\ref{conclusion_plot} gives an indication of the relative impact of the various datasets on the full GGM parameter space, where we marginalized over all NLSP types. In this figure the upper bound on the Higgsino mass is apparent by the presence of a meaningful limit from LEP, even for stop masses which are far outside of the reach of the experiment. For $M_{mess}=10^7$ GeV, the limits on the electroweakinos remain more important than the squark limits, even at the HL-LHC. The HL-LHC reach on the squark masses is therefore greater than one would expect from direct squark searches alone. This is no longer the case for high messenger scales ($M_{mess}=10^{15}$ GeV), where the direct squark bounds determine the reach of the HL-LHC.

Our analysis reveals a number of simplified topologies which, to the best of our knowledge, are currently not yet studied in detail. In particular, most attention so far has been devoted towards topologies with a bino NLSP, which are well-motivated benchmarks both because of their simplicity and because of their ubiquity in minimal gauge mediation. However now that the SUSY program at the LHC has matured, it may be worthwhile to investigate the topologies with Higgsino, wino or sneutrino NLSP somewhat more thoroughly. As we have shown in section \ref{sec:neutralNLSP}, their phenomenology can be very complex and interesting. It would therefore be useful to analyze these topologies in sufficient detail to establish accurate limits and/or to identify potential blind spots or possible improvements to the existing analysis strategies. For example, for the Higgsino/wino NLSP, the stop/sbottom can each decay to \emph{both} a $t$+MET and $b$+MET, which implies that stop/sbottom pair production can give rise to a $tb$+MET signature with a sizable branching ratio. Some of the latest CMS limits account now for this effect \cite{CMS:2016nhb,CMS:2016qtx}, and in particular for \cite{CMS:2016qtx} the sensitivity depends rather strongly on the branching ratios. Other interesting examples are inclusive searches for disappearing tracks optimized for colored production, or more generally simplified topologies with a wino and/or Higgsino and left-handed squarks (both 3th and lowest generations), and similar topologies with a sneutrino NLSP. 

\vspace{0.5cm}
\textbf{Acknowledgments:}\\
We are grateful to David Shih for the fruitful collaboration during the early stages of this work and for useful discussions later on. We further thank Patrick Draper, Alberto Mariotti, Michele Papucci, Filippo Sala and Giovanni Villadoro for useful discussions. We are grateful to Marcin Badziak, Lorenzo Calibbi, Alberto Mariotti, David Shih and Robert Ziegler for comments on the manuscript. The work of SK is supported by the LDRD Program of LBNL under U.S. Department of Energy Contract No.~DE-AC02-05CH11231. The work of DR is supported by the ERC grant Higgs at LHC.

\FloatBarrier

\appendix 

\section{NLSP life-time\label{app:NLSPdecay}}

A generic feature of gauge mediation scenarios is the presence of a light gravitino LSP whose couplings to the other MSSM particles are determined by the universal 2-body decay \cite{Martin:1997ns}
\begin{equation}
\Gamma ( \widetilde{X} \to X \tilde G )=\frac{m_{\widetilde{X}}^5}{48 \pi m_{3/2}^2 M_{Pl}^2} ~.\label{2body}
\end{equation}
The GGM phenomenology is then determined by the nature of the NLSP which decays to the gravitino LSP and its Standard Model partner. The NLSP life time in gauge mediation is then a function of the NLSP mass itself and the gravitino mass. The latter is directly related to the vacuum energy $\sqrt{F_0}$ via super-Higgs mechanism $m_{3/2}=\frac{F_0}{\sqrt{3}M_{P}}$. 

Given the soft spectrum at the messenger scale we can estimate the SUSY-breaking scale $F$ felt by the messengers by assuming a standard gauge mediation mechanism (with gaugino masses generated at 1-loop and scalar squared masses at 2-loops). Putting everything together we can write the decay length as 
\begin{equation}
l_{\tilde{X}}\approx \hbar c\, 16\pi \frac{F^2}{k^2 m_{\tilde{X}}^5},
\end{equation}
where the factor $k$ accounts for the fact that the vacuum energy can in general differ from the SUSY-breaking scale felt by the messenger sector (i.e $F= k F_0$ following the notation in \cite{Giudice:1998bp}). In calculable models of SUSY-breaking typically $k\lesssim1$, which we assume throughout this paper. However one should keep in mind that there are no general results putting an upper bound on $k$. 

Up to $\mathcal{O}(1)$ effects we can estimate the SUSY-breaking scale in terms of the gluino soft mass, which is typically the largest soft mass in our setup. We then get $F\approx  \frac{4\pi}{\alpha_3}M_{mess} m_{\tilde{g}}$ and rewrite the decay length of the NLSP as 
\begin{equation}
l_{\tilde{X}}\approx\frac{1.2\text{ m}}{k^2}\left(\frac{0.2\text{ TeV}}{ m_{\tilde{X}}}\right)^5\left(\frac{m_{\tilde{g}}}{5\text{ TeV}}\right)^2\left(\frac{M_{mess}}{10^7\text{ GeV}}\right)^2\ .
\end{equation}
This formula shows that our benchmark of $M_{mess}=10^{15}\text{ GeV}$ has long-lived NLSPs as a robust prediction independently on the details of the spectrum and on the particular UV completion. For $M_{mess}=10^{7}\text{ GeV}$ we are instead in intermediate regime where $\mathcal{O}(1)$ effects become important and the NLSP decay length will generically depend on the details of the model. In order to simplify the phenomenological discussion, we assume the NLSPs to be always long-lived, also for $M_{mess}=10^{7}\text{ GeV}$.

\section{The effect of compressed spectra}\label{app:squeezed}
\begin{figure}[h]\centering
\includegraphics[width=1\textwidth]{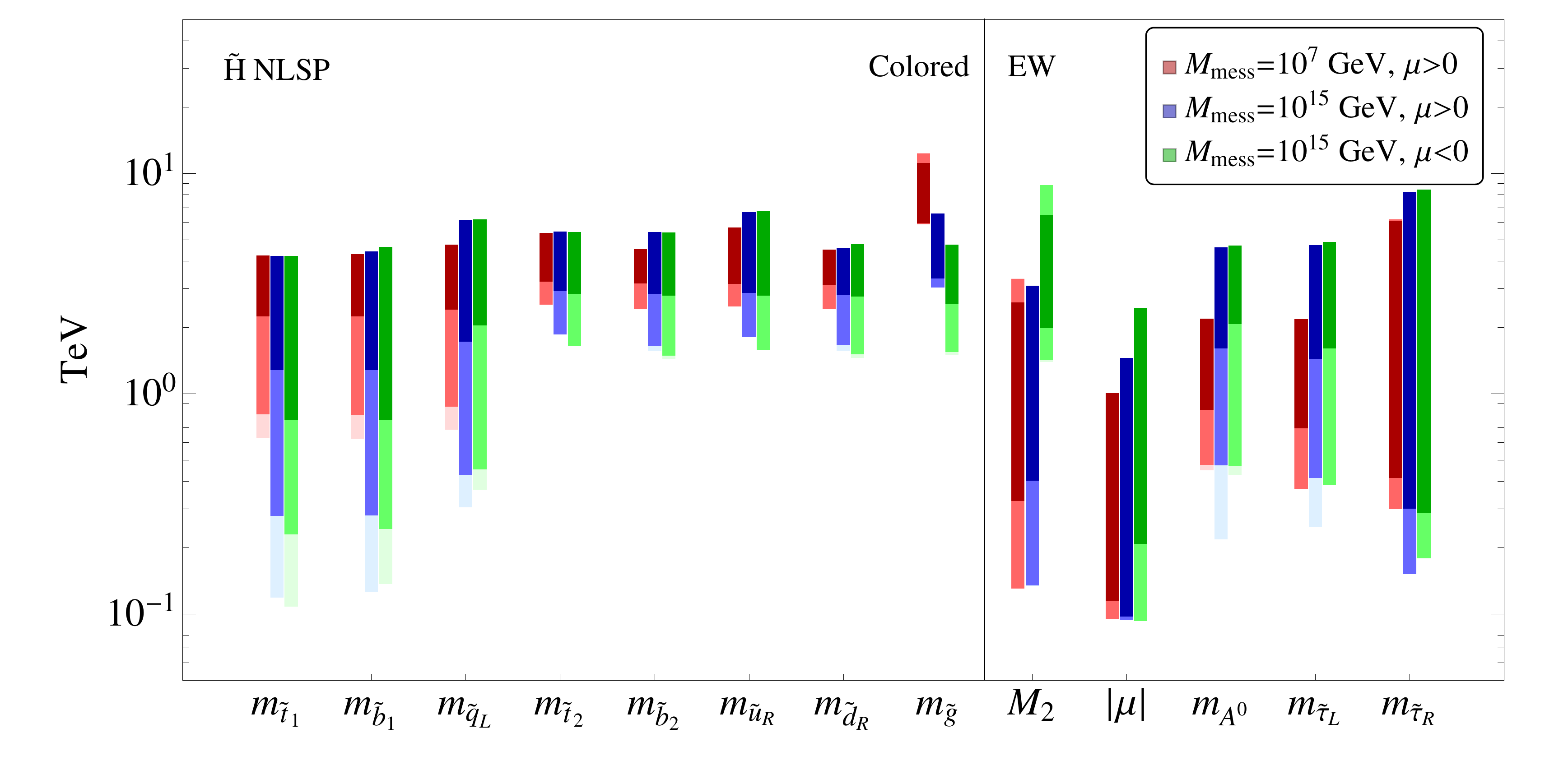}
\caption{Overview of the GGM parameter space for a Higgsino NLSP accounting for the deterioration of the  Regions with lightest shading are excluded or disfavored by existing data, regions with darker shading are accessible at the HL-LHC. There is no allowed parameter space for $M_{mess}=10^7$ GeV.}  \label{villadoroT}
\end{figure}
We briefly discuss here how the LHC reach on GGM scenarios is affected by accounted for compressed spectra, where we consider the Higgsino NLSP as a example. The wino NLSP is has similar features, while for the bino NLSP the NLSP mass uncorrelated with the remainder of the spectrum. Our results are for the Higgsino NLSP shown in fig.~\ref{villadoroT}, which should be compared with fig.~\ref{villadoroH} where compression was neglected. 

The lightest shading in fig.~\ref{villadoroT}, corresponding to the current bounds from the LHC, gets sensibly reduced with respect to the one in fig.~\ref{villadoroH}. The Higgsino NLSP can be compressed below the left-handed colored states, reducing the exclusion power of jet+MET, $2b+\text{MET}$ and $2t+\text{MET}$ searches. However it is important to keep in mind that, since we marginalize over the NLSP masses,  the importance of the spectrum configurations which minimize the LHC bounds gets magnified in fig.~\ref{villadoroT}. The exclusions presented here therefore correspond to the exceptions, rather than to ``typical'' spectra.

Notice however that we still get a lower bound on the scale of left-handed colored states in fig.~\ref{villadoroT}. That is because the presence of multiple colored states close in mass (the lightest stop, the lightest sbottom and the left-handed 1st and 2nd generation squarks) makes it difficult to squeeze the Higgsino NLSP below all of them making, such that some LHC searches remain effective. This is a common feature of SUSY scenario in which the production mechanism is dominated by left-handed states. 

Comparing the darker shaded bands between fig.~\ref{villadoroT} and fig.~\ref{villadoroH}, we see that compression effects are completely unimportant for HL-LHC and the asymptotic reach on GGM spectra at high-luminosity is essentially unchanged after accounting for compressed spectra. The reason is that the right-handed colored states become accessible and the upper bound on the mass of the Higgsino NLSP gives a minimum mass splitting between the right-handed states and the Higgsino itself. 
\FloatBarrier

\bibliography{biblio}

\end{document}